\documentclass[sn-standardnature]{sn-jnl}

\usepackage{fancyvrb}
\usepackage{hyperref}
\usepackage{natbib}
\usepackage{rotating}
\usepackage[resetlabels]{multibib}
\usepackage{placeins}

\newcommand\aap{Astron. Astrophys.}                
\newcommand\aj{AJ}                   
\newcommand\apj{Astrophys.~J.}                 
\newcommand\apjl{Astrophys.~J. Lett.}                
\newcommand\jai{J.~Astron. Instrum.} 
\newcommand\mnras{Mon. Not. R. Astron. Soc.}             
\newcommand\na{New~Astron.}          
\newcommand\nar{New~Astron.~Rev.}    
\newcommand\nastro{Nat.~Astron.}     
\newcommand\nat{Nature}              
\newcommand\prd{Phys. Rev.~D}        
\newcommand\prl{Phys. Rev.~Lett.}    
\newcommand\physrep{Phys.~Rep.}      
\newcommand\rnaas{Res. Notes AAS} 
\newcommand{\beq}{\begin{equation}}
\newcommand{\eeq}{\end{equation}}
\newcommand{\bea}{\begin{eqnarray}}
\newcommand{\eea}{\end{eqnarray}}

\newcommand{\movieDelTS}{1}
\newcommand{\movieDelTSWDM}{2}

\theoremstyle{thmstyleone}%
\theoremstyle{thmstyletwo}%

\theoremstyle{thmstylethree}%
\begin{document}

\title[The Dark Matter Clumping Signature in the 21-cm signal]{The Signature of Sub-galactic Dark Matter Clumping in the Global 21-cm Signal of Hydrogen}


\author[]{\fnm{Hyunbae} \sur{Park}$^{1,2,3}$}\email{hcosmosb@gmail.com}
\affil[]{$^1$ \orgdiv{Center for Computational Sciences}, \orgname{The University of Tsukuba},
\street{1 Chome-1-1 Tennodai}, \city{Tsukuba}, \state{Ibaraki} \postcode{305-8577}, \country{Japan}}

\affil[]{$^2$ \orgdiv{Computational Cosmology Center}, \orgname{Lawrence Berkeley National Laboratory},
\street{1 Cyclotron Road}, \city{Berkeley}, \state{California} \postcode{94720}, \country{USA}}

\affil[]{$^3$ \orgname{Kavli Institute for the Physics and Mathematics of the Universe}, \street{5-1-5 Kashiwanoha} \city{Kashiwa}, \postcode{Chiba 277-8583}, \country{Japan}}

\author[]{\fnm{Rennan} \sur{Barkana}$^{3,4}$}
\affil[]{$^4$\orgdiv{School of Physics and Astronomy}, \orgname{Tel Aviv University}, 
\city{Tel Aviv}, 
\postcode{69978}, \country{Israel}}

\author[]{\fnm{Naoki} \sur{Yoshida}$^{3,5,6}$}

\affil[]{$^5$\orgdiv{Department of Physics, School of Science}, \orgname{The University of Tokyo}, 
\street{7-3-1 Hongo},  
\city{Bunkyo}, \state{Tokyo} \postcode{113-0033}, \country{Japan}}

\affil[]{$^6$ 
\orgname{Max-Planck-Institut f\"{u}r Astrophysik}, \street{Karl-Schwarzschild-Str. 1} \city{Garching}, \postcode{D-85741}, \country{Germany}}

\author[]{\fnm{Sudipta} \sur{Sikder}$^4$}

\author[]{\fnm{Rajesh} \sur{Mondal}$^7$}

\affil[]{$^7$\orgdiv{Department of Physics}, \orgname{National Institute of Technology Calicut}, \city{Calicut}, \postcode{673601}, \state{Kerala}, \country{India}}

\author[]{\fnm{Anastasia} \sur{Fialkov}$^{8,9}$}

\affil[]{$^8$\orgdiv{Institute of Astronomy}, \orgname{University of Cambridge}, \street{Madingley Road} \city{Cambridge}, \postcode{CB3 0HA}, \country{UK}}

\affil[]{$^9$ \orgname{Kavli Institute for Cosmology}, \street{Madingley Road} \city{Cambridge}, \postcode{CB3 0HA}, \country{UK}}

\abstract{It is thought that the Universe went through an early period known as the Dark Ages, during which primeval density fluctuations grew to form the first luminous objects, marking the beginning of Cosmic Dawn around 100 million years after the Big Bang. The 21-cm line of hydrogen atoms is the most promising probe of these epochs, with extensive observational efforts underway. We combine hydrodynamical simulations with a large-scale grid in order to precisely calculate the effect of non-linear structure formation on the global (sky-averaged) 21-cm radio intensity. We show that it presents a potential opportunity to probe the properties of dark matter in a new regime, corresponding to a length-scale of only 150,000 light years and a mass-scale of 20 million Solar masses. This effect can in principle be detected unambiguously during the Dark Ages, where the weak signal requires an array of global signal antennae. During Cosmic Dawn, when stellar radiation boosts the signal, a single global antenna suffices, but the clumping effect must then be separated from the effect of the stars. Our findings open new avenues for testing the nature of dark matter as well as non-standard cosmological models.}

\maketitle

The standard model of cosmic structure formation posits that tiny density fluctuations generated in the very early Universe are amplified by the action of gravity. While the cosmic microwave background (CMB) radiation provides a direct view of the Universe at a cosmic age of 380,000 years, the evolutionary process during the subsequent Cosmic Dark Ages, as well as the formation of the first stars during Cosmic Dawn, remain observationally unexplored. The next moment that has been directly observed (by space-borne telescopes) is approximately 290 million years later \cite{2024Natur.633..318C}, when mature galaxies were already in existence. 

The most promising probe of the Dark Ages and Cosmic Dawn is the 21-cm line of hydrogen. The 21-cm intensity, measured as a brightness temperature, probes the spin temperature of hydrogen atoms, $T_{\rm S}$ — a measure of the population ratio between hyperfine energy states — which evolves through a complex interplay with the CMB temperature, $T_{\rm CMB}$, and the (kinetic) gas temperature, $T_{\rm gas}$. Unless we find previously unknown bright radio sources during these epochs \cite{2019MNRAS.483.1980M, 2024MNRAS.52710975S}, the 21-cm brightness temperature ($T_{\rm b}$) will be measured with respect to the CMB, which acts as a background source covering the entire sky.

At the highest redshifts, efficient collisional coupling kept $T_{\rm S}$ in equilibrium with $T_{\rm gas}$, producing a 21-cm absorption signal as the gas thermally decoupled from the CMB starting at $z \sim 200$. This is the expected 21-cm signal from the Dark Ages. The cosmic expansion lowered the collision rate over time, causing $T_{\rm S}$ to drift toward $T_{\rm CMB}$ from $z \sim 70$ and the 21-cm absorption signal to decline. Meanwhile, the baryons continued to cool more rapidly than the CMB through adiabatic expansion, reaching a temperature several times lower than the CMB at $z \sim 30$. 

As the first stars and galaxies began to form, stellar Lyman-$\alpha$ photons again induced a strong coupling between $T_{\rm S}$ and $T_{\rm gas}$. This time, the coupling occurred through the absorption and re-emission of Lyman-$\alpha$ photons that resulted in the Wouthuysen-Field effect \cite{1952AJ.....57R..31W,1959ApJ...129..536F,1959ApJ...129..551F}, leading to a large drop of $T_{\rm S}$ and another 21-cm absorption feature \cite{1997ApJ...475..429M}. Eventually the Lyman-$\alpha$ background grew strong enough to achieve saturated coupling, at which point $T_{\rm S}$ essentially equaled $T_{\rm gas}$.

The earliest time when there were enough stars for their radiation to substantially affect the average 21-cm signal marks a key milestone, which can be usefully defined as the end of the Dark Ages and the beginning of Cosmic Dawn. During the Dark Ages, 21-cm measurements can be used to place tight constraints on cosmological parameters \cite{Barkana2005, naoz05, 2023NatAs...7.1025M}, with the advantage of not needing to model complex astrophysical processes stemming from star formation. In the absence of stellar feedback, the matter in this era is expected to reflect the pristine conditions from the Big Bang.

During Cosmic Dawn, the complexities of astrophysics start to make the modeling and interpretation of the 21-cm signal more difficult. However, in broad surveys covering a wide range of possible astrophysical parameters of high-redshift galaxies, we showed that generically the first substantial astrophysical 21-cm effect is the above-mentioned Lyman-$\alpha$ coupling \cite{2017MNRAS.472.1915C, 2018MNRAS.478.2193C, 2021MNRAS.506.5479R}. The earliest cosmic heating, though highly uncertain, likely comes later, with cosmic reionization coming later still (see Methods). Thus, we expect an early stage of Cosmic Dawn during which the only important astrophysical effect is Lyman-$\alpha$ coupling from stellar radiation. This makes the 21-cm signal from this era potentially easier to model and interpret compared to that from later times.

Observationally, there are two primary approaches to detecting the 21-cm signal from early cosmic times. The first one is to use a radio interferometer to measure the spatial fluctuations in the 21-cm signal. Numerous ongoing collaborations are pursing this approach \cite{mwa_trott,lwa_garsden,nenufar,hera,lofar}. Great promise lies in the upcoming Square Kilometre Array (SKA) \cite{Koopmans}, with construction underway and scientific data expected in 5 years. 

Here we focus on the simpler approach of using a dipole antenna to observe the global 21-cm signal — that is, the mean radio intensity on the sky as a function of frequency (and thus redshift). There are many ongoing efforts to measure the global signal from Cosmic Dawn \cite{scihi2014,philip19,mist2024,de_lora2022}; the most significant results have been the tentative detection by the Experiment to Detect the Global EoR Signature (EDGES) \cite{2018Natur.555...67B} of a surprisingly deep absorption signal and the counter-evidence to this claim from the Shaped Antenna Measurement of the Background Radio Spectrum (SARAS) experiment \cite{SARAS3}. 

Focusing on the global signal is particularly reasonable in the case of the Dark Ages, where detecting the fluctuation signal would be a formidable task \cite{2023NatAs...7.1025M}. The wavelengths involved are so long that they are blocked by the Earth's ionosphere \cite{2021arXiv210305085B}. Consequently, there are many lunar or space-based experiments being developed as part of the international space race (see Supplementary Note Section~1).

Despite its importance, accurate modeling of the Dark Ages signal has received only limited attention. The signal is often calculated assuming a homogeneous Universe, disregarding the development of density structures. However, this approximation ignores the growth of non-linear structures at sub-Mpc scales that ultimately give rise to the first stars and galaxies. Another often overlooked factor is the baryon to dark matter relative streaming velocity, a residual spatially-variable velocity difference between dark matter and baryonic matter resulting from baryon acoustic oscillations in the early Universe \cite{2010PhRvD..82h3520T}. This streaming velocity affects the formation of the first galaxies in a non-uniform manner by delaying the growth of baryonic clumps, which can introduce 21-cm fluctuations at $\sim100$ Mpc scales during Cosmic Dawn \citep[e.g.,][]{2012Natur.487...70V,2019PhRvD.100f3538M}. This should similarly affect all non-linear density clumps.

Numerical simulations have shown that non-linear clumping increases the global 21-cm signal during the late Dark Ages \cite{2006NewAR..50..179A,2006ApJ...646..681S} and (assuming saturated Lyman-$\alpha$ coupling) reduces it during Cosmic Dawn \cite{2018ApJ...869...42X, 2021ApJ...923...98X}. However, these simulations did not account for the effect of the streaming velocity or large-scale density fluctuations, as they were limited to a box of size of 0.7~Mpc \cite{2006NewAR..50..179A,2006ApJ...646..681S} or $8~h^{-1}~{\rm Mpc}$ \cite{2021ApJ...923...98X}, where $h$ is the Hubble parameter defined as the Hubble constant devided by $100~{\rm km}~{\rm s}^{-1}~{\rm Mpc}^{-1}$. According to analytical calculations, the streaming velocity should induce a large-scale 21-cm fluctuation signal during the Dark Ages \cite{2014PhRvD..89h3506A}, and numerical simulations (that assumed saturated coupling) have shown a similar effect for the much later Epoch of Reionization \cite{2020ApJ...898..168C}. 

In this paper, we take a new look at how non-linear structure formation affects the global 21-cm signal during the Dark Ages and early Cosmic Dawn. We include the effect of the streaming velocity and also account for the large-scale density and velocity fields by combining small-scale hydrodynamic simulations with a large-scale semi-numerical simulation grid. We focus on two major implications. First, the effect of clumping is large enough to be potentially observable through the global signal, and second, such an observation would probe the strength of the primordial density fluctuations on remarkably small scales. This is an exciting prospect since various dark matter models beyond standard cold dark matter (CDM) produce small-scale cutoffs in the density power spectrum, including warm dark matter (WDM, in which the dark matter has a significant initial velocity dispersion) \cite{WDM} and fuzzy dark matter (FDM, in which an extremely low dark matter particle mass makes quantum effects appear on galactic scales) \cite{fuzzy}. 

Current direct observational measurements of the small-scale power spectrum reach (comoving) wavenumbers up to $k \sim 10$~Mpc$^{-1}$ through Lyman-$\alpha$ forest measurements, corresponding to a minimum WDM mass of $\sim 3$~keV \cite{WDMLya} and a minimum FDM mass of $\sim 2\times 10^{-21}$~eV \cite{FDM1,FDM2}, though systematic uncertainties from reionization and the astrophysical heating of the intergalactic medium remain a concern \cite{2021MNRAS.502.2356G}. A different constraint, derived from a combination of strong gravitational lenses and Milky Way satellite galaxies, may place a lower bound of 9.7~keV for WDM \cite{2021PhRvL.126i1101N}, corresponding to $k \sim 75$~Mpc$^{-1}$ \cite{2018MNRAS.481.1290M}.

By directly simulating cases with a power spectrum cutoff at small scales, we demonstrate below that the typical length scale probed by the global 21-cm signal is an order of magnitude smaller than current limits, corresponding to a mass scale that is 3 orders of magnitude smaller. Moreover, an important advantage of 21-cm observations is that they probe high redshifts, when the universe was more homogeneous and density fluctuations were more linear. For example, the density fluctuation in spheres of radius 0.1~Mpc reached unity at $z=6.2$, implying that this scale was only mildly non-linear at higher redshifts. At low redshifts, the combination of highly non-linear structure with astrophysical complexity makes the interpretation of cosmological measurements much more difficult.

While the probe of small scales through the global 21-cm signal is indirect, it is based on well understood fundamental physics that can be numerically simulated with high accuracy. This makes it a potentially powerful method to probe small scales and search for new clues to the nature of dark matter. There have been other ideas to use 21-cm measurements to constrain the small-scale power spectrum based on the abundance of low-mass galaxies \cite{2020PhRvD.101f3526M,2022PhRvD.106f3504F}. However, the complexities of star formation and astrophysical feedback (radiative, supernova, and black hole) make it likely impossible to achieve a convincing separation of the power spectrum cutoff effect on galaxy abundance from astrophysical effects. 

\section*{Results}

\begin{figure}[h]
\centering
\includegraphics[width=1.0\textwidth, angle=0]{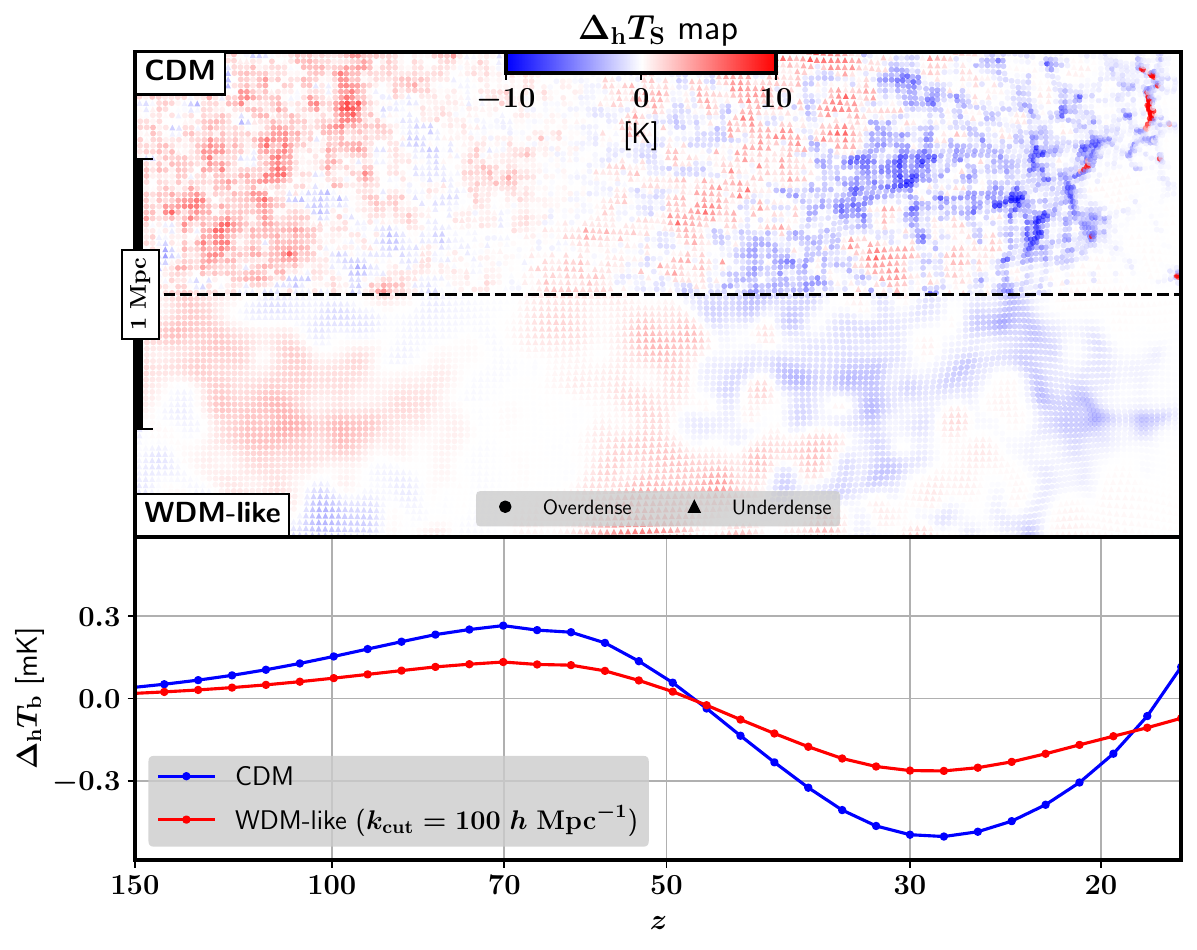} 
\caption{\textbf{Clumping effect in structure-formation simulations and resulting 21-cm signal.} \textbf{Top panel}: Map of gas particles, illustrating the spin temperature difference relative to the homogeneous Universe value, $\Delta_{\rm h} T_{\rm S}\equiv T_{\rm S}-T_{\rm S,h}$, for the CDM and WDM-like cases with $k_{\rm cut}=100~h~{\rm Mpc}^{-1}$, approximately corresponding to a dark matter particle mass of 7 keV. 
The map is a lookback image, with time elapsing toward the right, and the corresponding redshift labeled in the lower panel. Particles are represented as circular symbols in overdense regions and triangular symbols in underdense regions. $\Delta_{\rm h} T_{\rm S}$ is color-coded: red for positive values, white for zero, and blue for negative values (see also Supplementary Videos~\movieDelTS~and \movieDelTSWDM~for time-lapse videos of $\Delta_{\rm h} T_{\rm S}$ on a slice).
\textbf{Bottom panel}: Average difference in the 21 cm brightness temperature relative to the homogeneous Universe value, $\Delta_{\rm h} T_{\rm b}\equiv T_{\rm b}-T_{\rm b,h}$, for the CDM and WDM-like cases visualized in the upper panel, as a function of $z$. $\Delta_{\rm h} T_{\rm b}$ is sourced by $\Delta_{\rm h} T_{\rm S}$ depicted in the upper panel. We note that the case illustrated in this figure is for the simulation box at mean density and root-mean-square streaming velocity, while the comparison ``homogeneous" case refers here to the same simulation box but with a uniform density.} \label{fig_lookback}
\end{figure}

Figure~\ref{fig_lookback} illustrates how changes in the spin temperature due to structure formation during the Dark Ages affects the global 21-cm signal (see also Supplementary Note Section~2.2). On small scales, the complex relationship between density and spin temperature evolves with redshift, and the impact on the global signal depends on the non-linear asymmetry between overdensities and underdensities. The structure formation effect is suppressed by $\sim 50\%$ in the WDM-like scenario, where the initial conditions were smoothed with $k_{\rm cut}=100~h~{\rm Mpc}^{-1}$. We note that the results shown in this figure assume no astrophysical radiation (referred to as the Dark Ages case below). 

\begin{figure}[h]
\centering
\includegraphics[width=1.0\textwidth, angle=0]{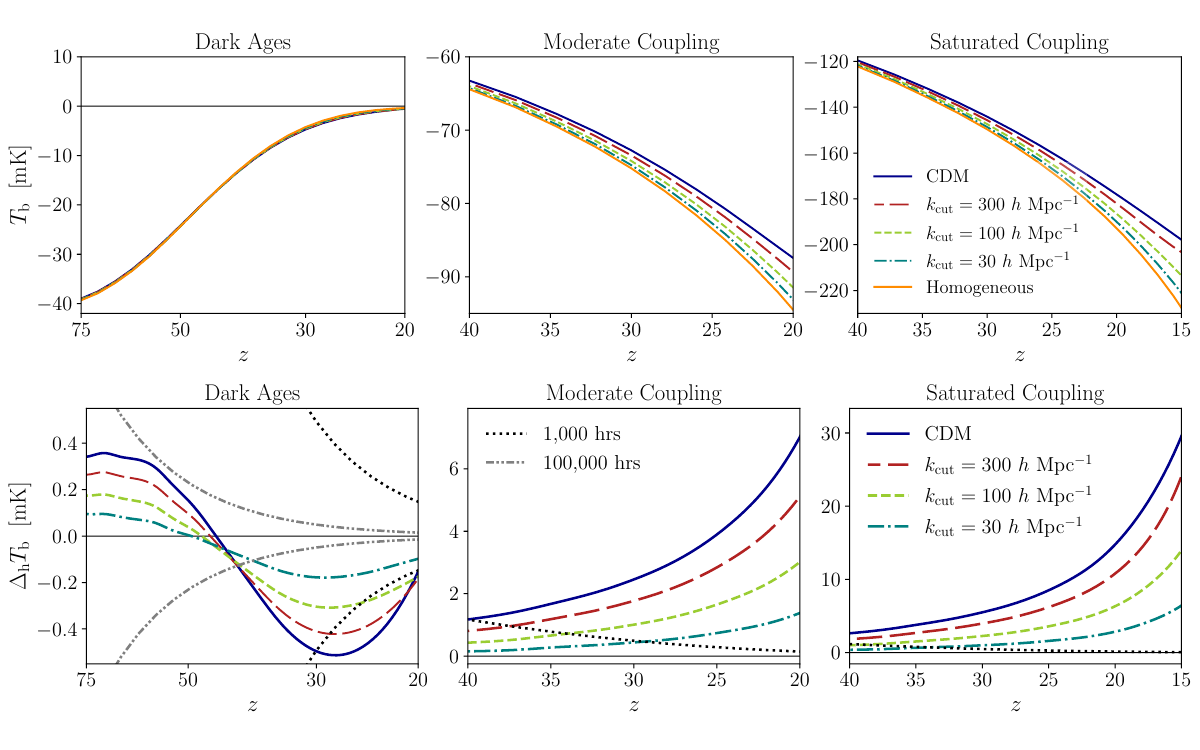}
\caption{\textbf{Predicted global 21-cm signal.} \textbf{Top panel:} The global 21-cm signal as a function of $z$, for a homogeneous universe, the standard CDM universe, and three cases of WDM-like models with a cutoff of the power spectrum on small scales. We consider three different levels of Lyman-$\alpha$ coupling, labeled Dark Ages, Moderate Coupling, and Saturated Coupling. \textbf{Bottom panel:} The effect of clumping, shown as the difference between the global signals in the various clumped cases with respect to the homogeneous case. Here we also show the expected thermal noise for a global signal experiment observing for integration times of 1,000\,hrs (black dotted curve) and (in the left panel) 100,000\,hrs (corresponding to an array of 100 global signal antennae, which we henceforth refer to as a global signal array; gray dashed-double-dotted curve); the thermal noise is shown for wide frequency ($\nu$) bins with $\Delta(\ln \nu) = 1$.} \label{fig_globalSignal}
\end{figure}

Figure~\ref{fig_globalSignal} presents the globally averaged 21-cm signal, $T_{\rm b}$, and the change in the signal due to structure formation ($\Delta_{\rm h} T_{\rm b} \equiv T_{\rm b} - T_{\rm b,h}$). We show the homogeneous Universe case, $T_{\rm b,h}$, as well as the final prediction for the standard CDM model, which includes the effect of density fluctuations on all scales (referred to here for brevity as ``clumping"). As explained above, we also include several WDM-like cases with different levels of smoothing ($k_{\rm cut} = 30$, 100, or 300 in units of $h~{\rm Mpc}^{-1}$). 

We show the results for three different levels of Lyman-$\alpha$ coupling, using the standard $x_\alpha$ parameter (which we originally introduced \cite{barkana05}) that describes the coupling strength. We label these cases the Dark Ages ($x_\alpha=0$, corresponding to the cosmological case of negligible astrophysical radiation), Saturated Coupling ($x_\alpha \rightarrow \infty$, the limiting case of an intensity of Lyman-$\alpha$ radiation that is high enough to fully couple the 21-cm spin temperature to the kinetic gas temperature), and Moderate Coupling ($x_\alpha=1$, an intermediate case where coupling is substantial but not close to saturated; this condition is a standard definition of the characteristic moment of the Lyman-$\alpha$ coupling transition). Note that we do not explicitly include star formation processes in our simulations; we only aim to illustrate the strength of the gas clumping effect in the presence of coupling, in order to motivate more detailed studies of its detectability in the presence of Lyman-$\alpha$ radiation from the first stars.

In standard CDM, in the Dark Ages case, clumping decreases the global signal (i.e., makes it less negative) at early times. The max positive $\Delta_{\rm h} T_{\rm b}=0.371$~mK at $z=64.4$. Moving forward in time, clumping then increases the signal after $z=45.6$, reaching a max negative $\Delta_{\rm h} T_{\rm b}=-0.510$~mK at $z=27.4$ before the clumping effect declines (along with the signal itself). The 21-cm signal of the WDM-like model with $k_{\rm cut} = 100\, h$~Mpc$^{-1}$ reaches a maximum difference from CDM of 0.206~mK at $z=27.2$.

For CDM at redshifts that may correspond to Cosmic Dawn ($z<40$), once stellar radiation provides substantial coupling, clumping always weakens the absorption signal, with $\Delta_{\rm h} T_{\rm b}$ increasing with time. For Moderate Coupling, $\Delta_{\rm h} T_{\rm b}=1.17, 2.46, 7.01$~mK at $z=40, 30$, and 20, respectively, while Saturated Coupling at the same redshifts gives $\Delta_{\rm h} T_{\rm b}=2.64, 5.50$, and 14.7~mK. The effect of clumping is positive in these Cosmic Dawn cases, since the presence of coupling makes the spin temperature of high density gas much higher than that of low density gas (with the latter's spin temperature being coupled and thus coming close to the low kinetic temperature of the same gas).

The Saturated Coupling case (along with the assumptions of negligible astrophysical heating and reionization) has a particular importance, as it gives the strongest absorption that is possible at each redshift in CDM (without invoking exotic models). Thus, this is an important baseline for comparison with current and future global 21-cm measurements. The strongest possible absorption in a homogeneous universe (the above naive case) at $z=40, 30, 25, 20, 17$, and 15, would be -122.2, -149.6, -168.3, -192.8, -211.9, and -227.3~mK, respectively. However, we find that structure formation in CDM reduces the maximum absorption at these redshifts to only -119.6, -144.1, -159.8, -178.0,  -189.7, and -197.8~mK, respectively. Thus, the effect of clumping increases with time, so that by redshift 15 it reduces the maximum possible global signal by 13.0\%. 

Figure~\ref{fig_globalSignal} also illustrates the observational thermal noise, for a 21-cm global signal antenna assuming a 1,000\,hr observation (a standard fiducial integration time, also equal to 11.4\% of a year). We show, in the Dark Ages case, the noise for the additional case of an effective integration time of 100,000\,hrs, potentially attainable with an array of global-signal antennae. Observations are more feasible at $z \lesssim 40$ (above 35 MHz), while higher redshifts have higher thermal noise and also likely require space or moon missions. Detecting the Dark Ages signal difference is difficult since the signal drops off as well towards redshift 20. However, if astrophysical radiation produces substantial coupling, this greatly increases the signal at Cosmic Dawn.

\subsection*{Observability}

Based on broad surveys covering a wide range of possible astrophysical parameters of high-redshift galaxies \cite{2017MNRAS.472.1915C, 2018MNRAS.478.2193C, 2021MNRAS.506.5479R}, we assume that the transition from the Dark Ages to Cosmic Dawn (defined as the moment of first substantial astrophysical influence on the 21-cm signal) occurs somewhere in the redshift range of $40-20$ (corresponding to a cosmic age of $65-180$ million years). We only consider observations at $z>20$, corresponding to the Dark Ages up to early Cosmic Dawn. To determine the observational significance of various detections, we follow our previous work \cite{2023NatAs...7.1025M,2024MNRAS.527.1461M}, and include thermal noise and optimistic foreground modeling, while neglecting uncertainties associated with calibration and beam modeling (see Supplementary Note Section~3).

Our full results are listed in Table~\ref{tab:obs}. We first consider the Dark Ages case. This has the advantage of the signal depending only on fundamental cosmology and thus being predictable with high accuracy. If this period extends down to lower redshift, this makes it easier to measure the signal and use it as an effective probe of cosmology and dark matter. The best case is having the entire range of $z=200-20$ available (where the observational power is concentrated below $z \sim 40$, as noted previously). In this case, with a global antenna (1,000~hr integration time), the global 21-cm signal in CDM can be detected (i.e., distinguished from no signal) with a statistical significance of 4.81$\sigma$, where $\sigma$ is the standard deviation of the measurement error. This compares to a homogeneous universe (as assumed in our previous work on detecting the Dark Ages signal \cite{2023NatAs...7.1025M}) which would yield a slightly higher 5.06$\sigma$. More importantly, the difference of the two (i.e., the effect of clumping) is detectable at 0.747$\sigma$. A global signal array (100,000~hr equivalent integration time) would increase these detection significances by a factor of 10, implying that the effect of clumping is potentially detectable at 7.47$\sigma$ significance.


\begin{table}
\begin{center}
\caption{{\bf The predicted observability of the global 21-cm signal.} We list the number of $\sigma$ achievable for: detecting (i.e., distinguishing from zero signal) the CDM signal; detecting the signal expected in a homogeneous universe; detecting the effect of clumping (i.e., the difference between CDM and homogeneous); and distinguishing the WDM-like case of $k_{\rm cut} = 100\, h$~Mpc$^{-1}$ from CDM or from homogeneous. We consider the Dark Ages, Moderate Coupling, or Saturated Coupling, over various redshift ranges. All results refer to a single global 21-cm antenna; multiply the number of $\sigma$ by 10 for a global 21-cm array.}
\label{tab:obs}

\begin{minipage}{\textwidth}
\begin{tabular*}
{\textwidth}{@{\extracolsep{\fill}}lccccc@{\extracolsep{\fill}}}
\toprule

{} & Detect & Detect & Detect & WDM from & WDM from \\

{} & CDM & homogeneous & clumping & CDM & homogeneous \\

\midrule

{\bf Dark Ages} & & & & & \\

{$z=200-20$} & 4.81 & 5.06 & 0.747 & 0.291 & 0.492 \\
{$z=200-30$} & 4.20 & 4.14 & 0.304 & 0.119 & 0.252 \\
{$z=200-40$} & 3.95 & 3.96 & 0.0568 & 0.0225 & 0.0385 \\

\midrule

{\bf Moderate} & & & & & \\

{$z=30-20$} & 74.0 & 81.3 & 7.39 & 4.19 & 3.20 \\
{$z=40-20$} & 110 & 119 & 9.52 & 5.41 & 4.11 \\

\midrule

{\bf Saturated} & & & & & \\

{$z=30-20$} & 152 & 168 & 15.4 & 8.74 & 6.65 \\ 
{$z=40-20$} & 226 & 246 & 20.0 & 11.4 & 8.61 \\

\botrule
\end{tabular*}

\end{minipage}
\end{center}
\end{table}


If the Dark Ages only go down to $z=30$, these numbers change to: 4.20$\sigma$ (for detecting the CDM signal with a global antenna, compared to 4.14$\sigma$ for the homogeneous universe), and 3.04$\sigma$ for detecting the effect of clumping with a global array. In the worst case of extending the Dark Ages only down to $z=40$, these numbers change to 3.95$\sigma$ (compared to 3.96$\sigma$), and 0.568$\sigma$ for detecting the effect of small-scale clumping. 

Thus, regardless of when the Dark Ages ended, by observing this era the global 21-cm signal can be detected at $4-5\sigma$ with a global signal antenna, and at $40-50\sigma$ with a global signal array. Detecting the effect of clumping requires a global signal array, and is possible only if the Dark Ages extend below $z \sim 30$. In the case of $z=20$, clumping can be detected at 7.47$\sigma$ significance, and the WDM-like case of $k_{\rm cut} = 100\, h$~Mpc$^{-1}$ can be distinguished from standard CDM at 2.91$\sigma$ and from the homogeneous case at 4.92$\sigma$.

If we consider Cosmic Dawn with substantial Lyman-$\alpha$ coupling of the 21-cm line to the gas temperature, then the 21-cm signal is much larger. This case is much easier to detect so we only consider here a single global antenna (1,000~hr integration time). As before, all the numbers below for detection significance would be 10 times higher for a global array. For WDM-like models, we only consider here the $k_{\rm cut} = 100\, h$~Mpc$^{-1}$ case.

The most pessimistic early Cosmic Dawn case we consider is Moderate Coupling from $z=30$ down to 20. In this case, the CDM signal can be detected at 74.0$\sigma$, and distinguished from homogeneous (i.e., the effect of clumping directly detected) at 7.39$\sigma$. The WDM-like model can be distinguished from CDM at 4.19$\sigma$ and from homogeneous at 3.20$\sigma$. If we instead assume Saturated Coupling over the same redshift range ($z=30-20$), the CDM signal can be detected at 152$\sigma$, with the effect of clumping detected at 15.4$\sigma$. WDM-like can be distinguished from CDM at 8.74$\sigma$ and from homogeneous at 6.65$\sigma$.

If we assume Moderate Coupling over a larger redshift range, $z=40-20$, then the CDM signal can be detected at 110$\sigma$, and distinguished from homogeneous at 9.52$\sigma$. WDM-like can be distinguished from CDM at 5.41$\sigma$ and from homogeneous at 4.11$\sigma$. Finally, our most optimistic early Cosmic Dawn case is Saturated Coupling over $z=40-20$. In this scenario, the CDM signal can be detected at 226$\sigma$, with the effect of clumping detected at 20.0$\sigma$. WDM-like can be distinguished from CDM at 11.4$\sigma$ and from homogeneous at 8.61$\sigma$.

\section*{Discussion}
\label{sec:conc}

We have calculated the global 21-cm signal expected during the Dark Ages and early Cosmic Dawn. By combining small-scale hydrodynamical simulations with a large-scale semi-numerical grid, we have shown that non-linear structure formation (clumping) has a substantial impact on the global mean. This effect is driven by density fluctuations on small scales, which presents a potential opportunity to probe the basic CDM model and the properties of dark matter in a new regime. Indeed, half the effect is due to (comoving) wavenumbers above $\sim 100\, h$~Mpc$^{-1}$,  corresponding to a radius (half a wavelength) of $\sim 50$~kpc, giving a characteristic halo mass scale (i.e., the total mass contained within such an initial comoving radius) of $2\times 10^7 M_{\odot}$. 

We can approximately translate this in the context of popular dark matter models that produce a small-scale cutoff by equating to where the perturbation amplitude (square root of the power spectrum) drops by a factor of 2 in these models. We find that 21-cm clumping would probe warm dark matter (WDM) \cite{WDM} masses around 7~keV and fuzzy dark matter (FDM) \cite{fuzzy} masses around $2\times 10^{-20}$~eV. 

We have shown that an array of global antennas could in principle detect the effect of clumping from the Dark Ages. This offers an exciting opportunity for a clear test of the popular Cold Dark Matter paradigm, which predicts substantial small-scale density fluctuations. During Cosmic Dawn, the 21-cm signal becomes much larger, and the clumping effect can in principle be detected with a single global antenna. In this regime, however, stellar radiation plays a role, and the clumping effect must be distinguished from it. We have illustrated the possible signal with simple examples of coupling strength, but in reality the coupling is expected to rise rapidly with time along with star formation. This different redshift dependence from the clumping will potentially facilitate their separation. 

We expect non-linear clumping to affect the 21-cm power spectrum as well, but we have focused here on the global 21-cm signal, which in principle requires a smaller effort to measure, though we have considered an optimistic scenario with respect to calibration and foreground removal. A large suite of upcoming experiments in the redshift range 20--40 (from Earth at the lower end, and space or the moon at the upper end) has a chance to search for the 21-cm signal that we predict in this study.

Finally, our findings open new avenues for testing non-standard cosmological models beyond WDM and FDM. Primordial magnetic fields (PMFs) or primordial black holes could amplify small-scale clumping \cite{B1,B2}. PMFs and dark matter–baryon interactions can also directly heat the intergalactic medium \cite{2024PhRvD.110l3506M,2015PhRvD..92h3528M}. Next-generation experiments offer exciting opportunities to uncover new insights into the nature of dark matter and other non-standard cosmological phenomena. Making 21-cm signal forecasts for these models will be essential to maximize the scientific return of the upcoming observations.

\section*{Methods}

\subsection*{Basic equations of the 21-cm signal}
\label{sec:equations}

We first briefly summarize here the main equations used for modeling the 21-cm signal. More details can be found in reviews or books on this subject \cite{2006PhR...433..181F, 2013fgu..book.....L, 2016PhR...645....1B, 2018enc1.book.....B, 2019cosm.book.....M}. We then briefly comment on the 21-cm effect of various types of astrophysical radiation.

In the intergalactic medium, the relative occupancy of the hyperfine spin states of hydrogen can deviate from thermal equilibrium with the CMB or with the normal (kinetic) gas temperature. In general, therefore, the relative occupancy is described in terms of the spin temperature $T_{\rm S}$ defined by
\bea
\frac{n_1}{n_0} = 3\exp{\left[-\frac{T_*}{T_{\rm S}}\right]}\ ,
\eea
where $T_*=E_{21}/k_{\rm B}=0.0682~{\rm K}$ is the energy of the 21-cm transition ($E_{21}=5.87\times10^{-6}~{\rm eV}$) divided by the Boltzmann constant, $k_{\rm B}=1.38\times10^{-17}~{\rm J}\cdot{\rm K}^{-1}$. 

Scattering with CMB photons drives $T_{\rm S}$ towards $T_{\rm CMB}$, but atomic collisions and the Wouthuysen-Field effect \cite{1952AJ.....57R..31W, 1959ApJ...129..536F,1959ApJ...129..551F} draw $T_{\rm S}$ toward $T_{\rm gas}$ as described by
\bea
T_{\rm S}^{-1} = \frac{T_{\rm CMB}^{-1}+x_{\rm tot} T_{\rm gas}^{-1}}{1+x_{\rm tot}}\ .
\eea
Here, $x_{\rm tot}=x_\alpha+x_{\rm c}$ is the sum of the Lyman-alpha (Ly$\alpha$) coupling coefficient ($x_\alpha$) and collisional coupling coefficient ($x_{\rm c}$). In the Dark Ages, there is no Ly$\alpha$ coupling ($x_\alpha= 0$), giving $x_{\rm tot}=x_{\rm c}$. During Cosmic Dawn, Ly$\alpha$ coupling is introduced by star-forming galaxies producing ultraviolet light. In the limit of strong Ly$\alpha$ coupling ($x_\alpha\rightarrow \infty$), the spin temperature is perfectly coupled to the gas temperature (i.e., $T_{\rm S}=T_{\rm gas}$). We include the additional low-temperature effects that result in an effective $T_{\rm gas}$-dependent reduction in the Ly$\alpha$ coupling, and can be found in the above general reviews. 

As we noted in the main text, in broad surveys covering a wide range of possible astrophysical parameters of high-redshift galaxies, we showed \cite{2017MNRAS.472.1915C, 2018MNRAS.478.2193C, 2021MNRAS.506.5479R} that generically the first substantial astrophysical effect is Ly$\alpha$ coupling. Thus, in this work we illustrate the possible signal from this epoch (which might be called early Cosmic Dawn) by including different levels of Ly$\alpha$ coupling $x_a$, as we describe in the main text and in Supplementary Note Section~2.3. 

The 21-cm signal is sourced by the difference between $T_{\rm S}$ and $T_{\rm CMB}$, giving the brightness temperature $T_{\rm b} = (1+z)^{-1}(T_{\rm S}-T_{\rm CMB})(1-e^{-\tau})$, where $\tau$ is the optical depth at 21~cm:
\bea \label{eq_tau}
\tau(z) = \frac{3c\,\lambda^2_{21}h_{\rm P} A_{10} \,n_{\rm HI}}{32\pi k_{\rm B} \, T_{\rm S} (1+z)(dv_r/dr)}\ .
\eea
Here, $\lambda_{21}=21~{\rm cm}$, $h_{\rm P}=6.626\times10^{-27}~{\rm erg}\cdot{\rm s}$ is the Planck constant, $A_{10}=2.85\times 10^{-15}~{\rm s}^{-1}$ is the spontaneous decay rate of the hyperfine transition, $n_{\rm HI}$ is the neutral hydrogen number density, and $dv_r/dr$ is the velocity gradient along the line of sight. 

The net $T_{\rm b}$ in a simulation volume is obtained by volume-averaging the signal over resolution elements. In our simulations, where the particle mass is fixed, the average is given by
\bea \label{eq_averaging} 
\left< T_{\rm b}\right> = \frac{1}{N_{\rm ptl}} \sum_{i=1}^{N_{\rm ptl}} W_{V,i} T_{{\rm b},i}, 
\eea
where the volume weight for each particle, $W_{V,i}$, is calculated by dividing the cosmic mean density, $\bar\rho_M$, by the particle density, $\rho_{M,i}$. This can be thought of as the correction from our Lagrangian code to the (actual, physical) Eulerian volume.

Here we elaborate a few details on astrophysical radiation that becomes important at later times. One important process is cosmic heating, which is mainly due to X-ray sources from stellar remnants, plus weak heating associated with the Ly$\alpha$ photons themselves, and another is cosmic reionization due to stellar ultra-violet photons. Cosmic reionization is constrained by CMB measurements \cite{planck:2018}, and in most models it becomes substantial well below redshift 15. The most uncertain effect is cosmic heating. The properties of high-redshift X-ray sources such as their intensity and spectrum are largely unconstrained. Still, even in models that vary these properties widely, substantial X-ray heating usually occurs below redshift 20, and later than the rise of Ly$\alpha$ coupling that is expected at redshift 30 or below. Ly$\alpha$ coupling is itself accompanied by Ly$\alpha$ heating by the same photons \cite{1997ApJ...475..429M, 2004ApJ...602....1C, 2006ApJ...651....1C, 2007ApJ...655..843C, 2006MNRAS.372.1093F}, but this heating becomes important only when the coupling is already quite strong ($x_{\alpha} > 10$) \cite{2021MNRAS.506.5479R}. Thus, the Saturated Coupling case that we consider here, without Ly$\alpha$ heating, serves as an upper limit on the signal. It has also been suggested that there is an additional heating mechanism through the interaction with the CMB \cite{2018PhRvD..98j3513V} but this is not expected to produce a meaningful effect \cite{2021RNAAS...5..126M}.

\subsection*{Numerical simulations} \label{sec:calculation}

Baryons form highly non-linear structures toward the end of the Cosmic Dark Ages, necessitating numerical simulations for accurate calculation of the 21-cm signal. We employ the Smoothed Particle Hydrodynamics (SPH) simulation code GADGET \cite{2006ApJ...652....6Y,2007ApJ...663..687Y}, modified to incorporate chemical reactions involving hydrogen and helium species, as well as Compton heating. Compton heating counteracts the adiabatic cooling of gas caused by cosmic expansion by transferring energy from CMB photons to baryons during the Dark Ages  \cite{2001NewA....6...79S,2005MNRAS.364.1105S}, mediated by residual free electrons that froze out during cosmic recombination. The number density of these electrons evolves gradually as they continue to recombine with hydrogen and helium ions. All these processes are fully coupled with the gravity-hydrodynamical calculations in our code.

We initialize the simulation in a 3~Mpc cubic box containing $512^3$ dark matter and baryon particles, with masses of $m_{\rm DM} = 6670~M_\odot$ and $m_{\rm gas} = 1250~M_\odot$, respectively, for the SPH calculation to reproduce structure formation at sub-galactic scales. The fixed mass resolution of SPH is well-suited for modeling gas physics in dense structures, where higher spatial resolution is required compared to low-density regions. SPH is known to be inherently limited in accurately describing abrupt discontinuities in physical quantities resulting from strong shocks. However, such strong shocks are rare in our simulations of the Dark Ages and early Cosmic Dawn. Mild shocks arising from streaming velocities and the growth of structure in these epochs have been shown to be in good agreement between Lagrangian and Eulerian simulations \cite{2012ApJ...760....4O}.

Obtaining the globally averaged signal requires accounting for variations in structure formation rates beyond the scale of our simulation box. Specifically, the local density in a 3 Mpc box can deviate substantially from the cosmic mean, leading to accelerated structure formation at overdensities and slower formation at underdensities \cite[e.g.,][]{2020ApJ...900...30P}. Additionally, the relative streaming velocity between baryons and dark matter, which fluctuates over 100 Mpc scales (and thus can be approximated as uniform in a 3~Mpc box) \cite{2010PhRvD..82h3520T}, can delay the growth of baryonic structures at sub-galactic scales \cite{2011ApJ...736..147G,2012ApJ...760....4O,2013ApJ...763...27N,2013ApJ...771...81R,2016PhRvD..93b3518A,2019MNRAS.484.3510S}. 

We generate initial conditions at $z=200$ with 17 different combinations of local overdensity ($\bar\delta$) and streaming velocity ($V_{bc}$) to obtain the functional dependence of the 21-cm signal on these parameters. The magnitude of the streaming velocity decays as $1 + z$ and can be described by its $z = 1000$ value, denoted as $V_{bc,i}$. The distribution of $V_{bc,i}$ follows a Boltzmann distribution which peaks near $28~{\rm km}~{\rm s}^{-1}$. Fifteen cases are derived from all possible combinations of $V_{bc, i} = 0$, 28, and 56 ${\rm km}~{\rm s}^{-1}$ and $\bar{\delta}/\sigma_* = -2, -1, 0, 1,$ and 2, where $\sigma_* \equiv 0.014$, which is approximately the standard deviation of the density in 3 Mpc boxes at $z = z_i=200$. The remaining two cases are for $\bar{\delta}/\sigma_* = -3$ and 3 with $V_{bc, i} = 28~{\rm km}~{\rm s}^{-1}$. Supplementary Figure~1 provides an overview of the simulation across different choices of $\bar\delta$ and $V_{bc,i}$ in the CDM scenario. Note that the effect of varying $\bar\delta$ is large and easily noticeable, while that of varying $V_{bc,i}$ is more subtle. The smoothing effect due to streaming is more pronounced in the outskirts of the visualized region, where less massive structures dominate.

In addition to the base set of 17 simulations that assumes cold dark matter with a standard initial power spectrum of density fluctuations, we run three additional sets with varying levels of Gaussian smoothing on the initial density and velocity fields, applying $k_{\rm cut}=30,100$, or $300~h~{\rm Mpc}^{-1}$, giving additional 51 simulations. In these additional sets, we multiply each of the Fourier modes in the initial conditions by $\exp(-k^2/k^2_{\rm cut})$, to damp the fluctuation modes at wavenumber $k \gtrsim k_{\rm cut}$. As noted above, such a cutoff provides a useful generic model that can mimic the effects of various non-standard dark matter models, such as WDM or FDM (although these particular models produce additional dynamical effects that may somewhat modify the results); we refer to these models as ``WDM-like".

We deactivate cooling by hydrogen atoms and molecules to prevent the collapse of gas in minihalos and obtain $T_{\rm gas}$ in the limit of no star formation. The collapse typically begins in minihalos with masses of $10^6~M_\odot$, starting at $z \sim 30$. However, even after the first stars form, heating from star formation activities should affect only a small fraction of baryons in the Universe. Therefore, we consider the $T_{\rm gas}$ statistics in our simulation valid down to early Cosmic Dawn ($z\approx 20$). Ly$\alpha$ radiation comes from distances that are typically much larger than our box size, so we illustrate cases where its effect is substantial, with Moderate Coupling ($x_\alpha=1$) or Saturated Coupling ($x_\alpha=\infty$) as described in the main text. 

The cosmological parameters adopted are a total matter density of $\Omega_{\rm M}=0.3111$, baryonic matter density of $\Omega_{\rm b}=0.0490$, and Hubble parameter of $h=0.6766$ \cite{planck:2018}. All quoted lengths are in comoving units. 

\subsection*{The homogeneous Universe signal} \label{sec:uniform}

A major challenge of this work is to calculate accurately small deviations in the global averages of $T_{\rm S}$ and $T_{\rm b}$ from their homogeneous Universe values, $T_{\rm S,h}$ and $T_{\rm b,h}$, due to structure formation. To make sure that we are isolating the effect of clumping and not confusing it with numerical errors, we obtain the homogeneous Universe case also from our simulations, instead of from an external comparison case. Specifically, we run ``uniform" simulations, where all fluctuations in the initial conditions (ICs) are removed by setting the displacement and velocity of every particle to zero and assigning the average gas temperature to all particles. We generate a uniform counterpart for every combination of $\bar{\delta}$ and $V_{bc,i}$ with $2\times 128^3$ particles in the same simulation box (3 Mpc), resulting in 17 uniform cases. These uniform simulations are used for testing the accuracy of the chemical processes implemented in our simulation, and controlling the numerical errors in $T_{\rm S}$ and $T_{\rm b}$.

The standard CAMB code outputs the free electron fraction with respect to nuclei ($x_e\equiv n_e/n_{\rm nuc}$), $T_{\rm gas}$, $T_{\rm S}$, and $T_{\rm b}$ calculated for the homogeneous Universe in the Dark Ages regime (i.e., $x_\alpha=0$) \cite{Lewis2007,CAMB}, which should match results from our uniform simulation at the cosmic mean density (with no streaming velocity). In Supplementary Figure~2, we show the fractional error of these quantities against CAMB results for $z$ between 200 and 30. The fractional error is at most $\sim 1\%$ for those quantities. Notably, the fractional error in $T_{\rm S, h}$ is much smaller than 1\% throughout the redshift range, but $T_{\rm b}$, which is derived from $T_{\rm S}$, shows a higher error fraction of $\sim 1\%$. This is because $T_{\rm b}$ is proportional to $T_{\rm S}-T_{\rm CMB}$, where the fractional error is amplified at lower redshifts as $T_{\rm S}$ approaches $T_{\rm CMB}$ due to weakening collisional coupling. This suggests that the accuracy of our results is even better for the cases with substantial Ly$\alpha$ coupling.

The error in $T_{\rm S}$ and $T_{\rm b}$ arises from GADGET ignoring the radiation component in simulating cosmic expansion and from accumulated integration errors due to finite time steps. We find that these errors introduce substantial noise and offset in $\Delta_{\rm h} T_{\rm b}$ at $z\gtrsim 50$, where $\Delta_{\rm h} T_{\rm b}$ is below $1\%$ of $T_{\rm b}$. Most cosmological simulation codes, including GADGET, are not optimized for sub-percent accuracy in these quantities, as such precision is generally unnecessary for typical applications. However, here we do need to reach a percent level accuracy because the net $\Delta_{\rm h} T_{\rm b}$ results from a small, percent-level imbalance between the signals from overdense and underdense regions, as we demonstrate in Supplementary Note. The uniform simulations are thus forced to follow exactly the same time steps down to $z\approx50$ as their non-uniform counterparts.

By taking the uniform simulation results (instead of the CAMB results) as the comparison homogeneous Universe values, $T_{\rm S,h}$ and $T_{\rm b,h}$, we expect to largely cancel out the errors in overall normalization when we calculate the effect of structure formation. Additionally, both the uniform and main simulations calculate $T_{\rm gas}$ at the cosmic mean without the radiation component in the cosmic expansion, thereby matching the normalization errors at least to the leading order in $T_{\rm S}$ and $T_{\rm b}$. This approach enables us to achieve a high accuracy in calculating the {\it differences} due to structure formation, $\Delta_{\rm h} T_{\rm S} = T_{\rm S} - T_{\rm S,h}$ and $\Delta_{\rm h} T_{\rm b}  = T_{\rm b} - T_{\rm b,h}$. Another consistency check can be seen in Supplementary Note Section~2.2, where we find that the simulated $T_{\rm S}$-density relation intersects with $T_{\rm S,h}$ precisely at the cosmic mean density at each redshift.

\subsection*{Resolution convergence} \label{sec:tests}

To validate our fiducial resolution of $\mathcal{N} = 512$ particles on a side in a $3~{\rm Mpc}$ box, we compare simulation results from $\mathcal{N} = 64,~128,~256,~512,$ and $1024$ in the same box size, where the total number of particles is given by $2\mathcal{N}^3$. For this comparison, we set the local density and streaming velocity to the cosmic mean ($\bar{\delta}=0$) and the typical value ($V_{bc,i}=28~{\rm km}~{\rm s}^{-1}$), respectively. We show a visual comparison of these simulations in Supplementary Figure~3.

Supplementary Figure~4 shows the change in $T_{\rm b}$ due to structure formation, $\Delta_{\rm h} T_{\rm b}$, for the five resolutions. The results indicate that the differences due to increasing resolution are generally much smaller than the main signal caused by cosmic structure growth and decrease rapidly with increasing resolution--suggesting that the results are well-converged. For instance, $\Delta_{\rm h} T_{\rm b}$ at $z=30$ for $\mathcal{N} = 64$, 128, 256, 512, and 1024 are -0.291, -0.371, -0.448, -0.497, and -0.517 mK in the Dark Ages limit, and 2.87, 3.47, 4.09, 4.53, and 4.80 mK in the Saturated Coupling limit. The fractional differences in $\Delta_{\rm h} T_{\rm b}$ when increasing $\mathcal{N}$ from 64 to 128, from 128 to 256, from 256 to 512, and from 512 to 1024 are 28\%, 21\%, 11\%, and 4\% in the Dark Ages limit, and 21\%, 18\%, 11\%, and 6.7\% in the Saturated Coupling limit at the same redshift, respectively. 

\subsection*{The global 21-cm signal}
In order to accurately calculate the globally averaged 21-cm brightness temperature, we must go beyond the small (3~Mpc) simulated volumes, as each is not representative of the overall cosmic average. The local average 21-cm signal in each simulation box depends on the average density of the box, and the large-scale density fluctuations do not average out exactly due to the non-linear dependence of the 21-cm signal on the local density. A further effect comes from the baryon to dark matter streaming velocity and its variation on large spatial scales. 

Our method thus combines the small-scale hydrodynamical simulations with a large-scale realization of the Universe in a cubic volume that is 384~Mpc on a side using our 21-cm Semi-numerical Predictions Across Cosmic Epochs (\texttt{21\textsc{cm}SPACE}) \cite{2012Natur.487...70V,2014Natur.506..197F} simulation code. The density and baryon-dark matter streaming velocity fields are generated on a $128^3$ mesh so that the 3~Mpc pixel size matches our hydrodynamical simulation box size. By interpolating (and slightly extrapolating, all using cubic spline) the tabulated 21-cm signal for 17 combinations of density and drift velocity in $(3~{\rm Mpc})^3$ volumes described above, we calculate the 21-cm signal $\Delta_{\rm h} T_{\rm b}$ for every pixel of the large-scale grid, obtaining the globally averaged 21-cm signal while accounting for both large-scale fluctuations (from 3 up to several hundred Mpc scales) and non-linear structure formation on sub-Mpc scales.

\section*{Data availability}
Data points for the lower panel of Figure~\ref{fig_lookback} are shared in Source Data. 
Intermediate data products such as snapshots of three-dimensional hydrodynamical simulations are available upon request to the corresponding author.

\section*{Code availability}
CAMB is available at \url{http://camb.info}. The BCCOMICS package is available at \url{https://github.com/KJ-Ahn/BCCOMICS}.
The GADGET code used for this work is available upon request to the corresponding author. 

\section*{Acknowledgments}
HP thanks Kana Moriwaki, Yajima Hidenobu, and Kyungjin Ahn for helpful comments on this work. HP also thanks Jiten Dhandha for providing his simulation data. HP was supported in part by grant NSF PHY-2309135 to the Kavli Institute for Theoretical Physics (KITP). Numerical simulations for this work were performed on the idark computing cluster of the Kavli Institute for Physics and Mathematics of the Universe, the University of Tokyo. RB and SS acknowledge the support of the Israel Science Foundation (grants no.\ 2359/20 and 1078/24). RM is supported by the NITC FRG Seed Grant (NITC/PRJ/PHY/2024-25/FRG/12). NY acknowledges financial support from JSPS International Leading Research 23K20035. RB and NY acknowledge JSPS Invitational Fellowship S24099.

\section*{Author contributions}
RB initiated the project. NY and HP developed the numerical codes. HP ran the simulations, wrote the text, and produced most of the figures in consultation with RB and NY. SS interpolated the simulation results, ran the large-scale grid, and made Figure~2, in consultation with RB and AF, and using the \texttt{21\textsc{cm}SPACE} code originally developed by AF and RB. RM calculated the observational detection significance. All the authors edited the text.

\section*{Competing interests}
The authors declare no competing interests. 
\FloatBarrier
\FloatBarrier


\end{document}


\title[The Dark Matter Clumping Signature in the 21-cm signal]{Supplementary Note: The Signature of Sub-galactic Dark Matter Clumping in the Global 21-cm Signal of Hydrogen}

\maketitle

In this Supplementary Note, we first list various Dark Ages experiments (Sec.~\ref{sec:Dark}). We then discuss simulation results on the impact of the complex baryonic and structure-formation physics on the global 21-cm signal of hydrogen (Sec.~\ref{sec:complex_physics}). Finally, we give more details on how we model the observability and the partial degeneracy of the signal with the foreground (Sec.~\ref{sec:observe}). 

\section{Information on Dark Ages experiments} \label{sec:Dark}

We provide here more detailed information on ongoing and planned observations mentioned in the introduction in the main text. With respect to the global signal from the Dark Ages, there are many lunar or space-based experiments being developed around the world as part of the international space race, in the USA, China, India, Europe, and Japan. They include DAPPER ({\url{https://www.colorado.edu/project/dark-ages-polarimeter-pathfinder}}) (USA), FARSIDE ({\url{https://www.colorado.edu/project/lunar-farside}}) (USA), DSL/Hongmeng ({\url{https://www.astron.nl/dsl2015}}) (China), PRATUSH ({\url{https://wwws.rri.res.in/DISTORTION/pratush.html}}) (India), FarView ({\url{https://www.colorado.edu/ness/projects/farview-lunar-far-side-radio-observatory}}) (USA), SEAMS (India), LuSee Night ({\url{https://www.lusee-night.org/night}}) (USA), ALO ({\url{https://www.astron.nl/dailyimage/main.php?date=20220131}}) (Europe), ROLSES ({\url{https://www.colorado.edu/ness/projects/radiowave-observations-lunar-surface-photoelectron-sheath-rolses}}) (USA), and Tsukuyomi ({\url{https://www.sankei.com/article/20240517-5ZRIGLFVXJP4RA4CRVHJBTDED4/}}) (Japan).

\begin{sidewaysfigure}[h]
\centering
\includegraphics[width=\textwidth, angle=0]{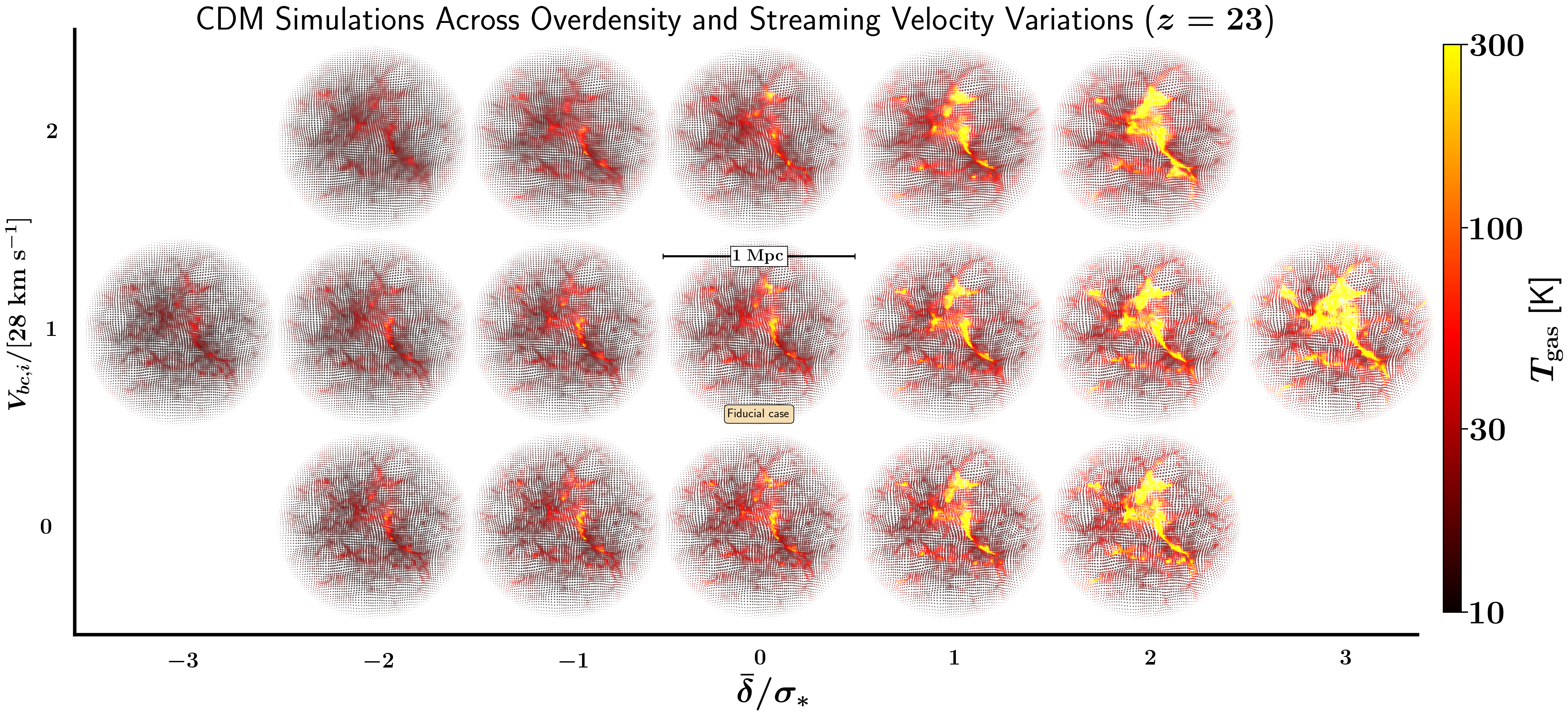}
\caption{Gas particles in 17 CDM simulations with varying box overdensity and baryon–dark matter streaming velocity. A circular region with a 1 Mpc diameter, centered on the largest structure at redshift $z=23$ is shown for simulations with different parameter choices. The gas temperature $T_{\rm gas}$ of each particle is color-coded such that black, red, and yellow approximately correspond to 10, 50, and 300 K, respectively. Marker size decreases toward the edges to highlight the central structure. A full description for the normalized overdensity ($\bar\delta/\sigma_*$) and the streaming velocity ($V_{bc,i}$) is provided in the Numerical simulations section under Methods. Panels are arranged such that $\bar\delta$ increases from $-3\sigma_*$ to $3\sigma_*$ from left to right, and $V_{bc,i}$ increases from 0 to $56~{\rm km}~{\rm s}^{-1}$ from bottom to top. } \label{fig_Ptl}
\end{sidewaysfigure}

\begin{figure}[h]
\centering
\includegraphics[width=1\textwidth, angle=0]{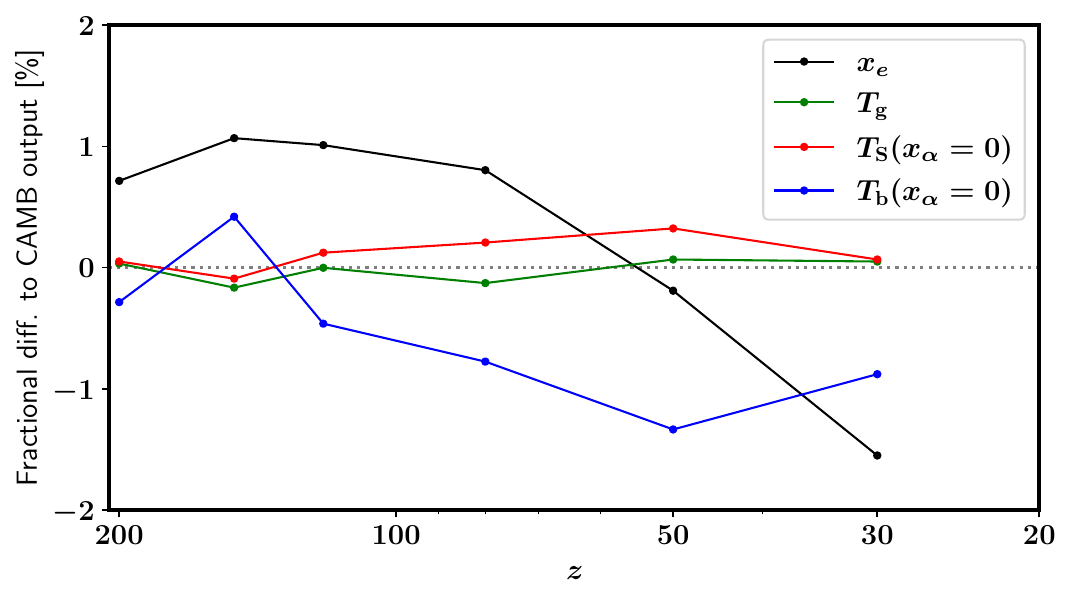} 
\caption{ Fractional differences in physical quantities between the uniform simulation and CAMB. The differences are shown as a function of redshift for the free electron fraction ($x_e\equiv n_e/n_p$), $T_{\rm gas}$, $T_{\rm S}$, and $T_{\rm b}$. This comparison illustrates the accuracy of our simulation in calculating quantities related to the global 21-cm signal.} \label{fig_compare2CAMB}
\end{figure}

\begin{figure}[h]
\centering
\includegraphics[width=1\textwidth, angle=0]{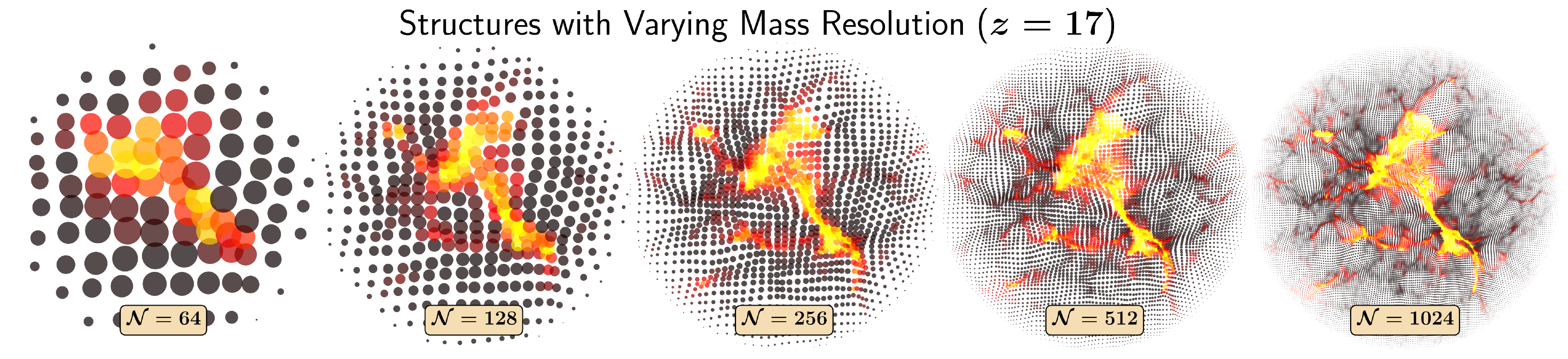} 
\caption{Gas particles in simulations with various mass resolutions. Snapshots at $z = 17$ are shown for simulations with five mass resolutions, corresponding to $\mathcal{N} = 64$, 128, 256, 512, and 1024 particles per side, using the fiducial box density ($\bar\delta = 0$) and streaming velocity ($V_{bc,i} = 28~{\rm km}~{\rm s}^{-1}$). Marker sizes are inversely proportional to $\mathcal{N}$. Other figure details are as in Supplementary Figure~\ref{fig_Ptl}.} \label{fig_ResTestPtl}
\end{figure}

\begin{figure}[h]
\centering
\includegraphics[width=1\textwidth, angle=0]{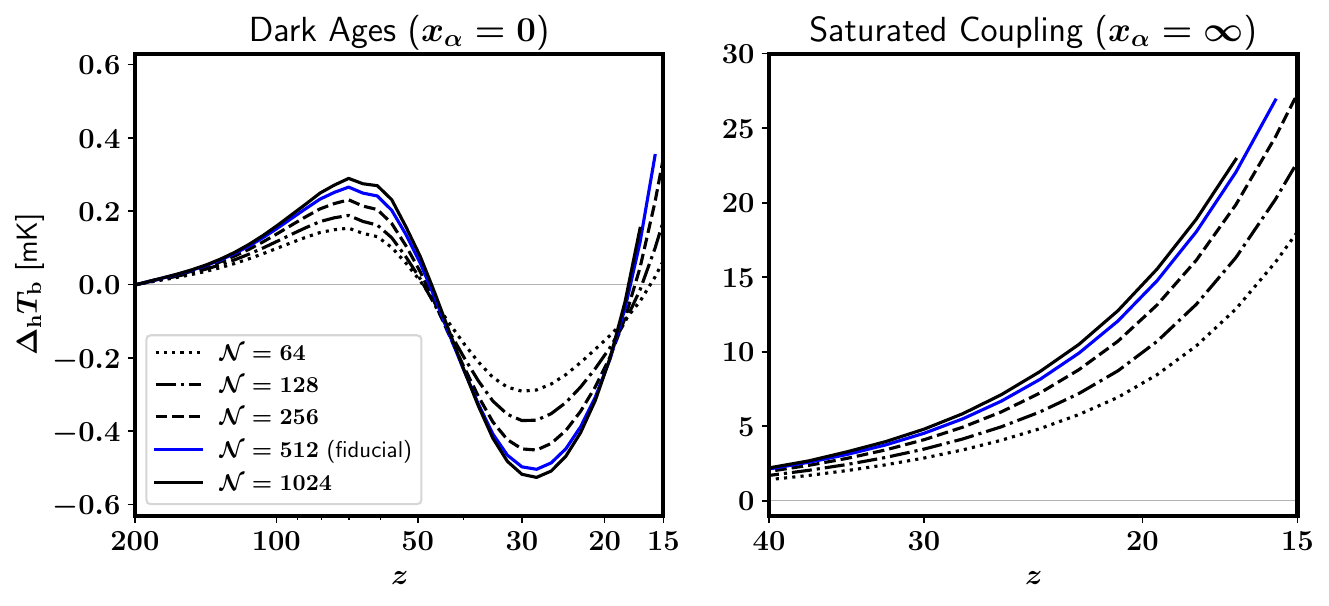} 
\caption{ Clumping signal in the global 21-cm background with various mass resolutions. $\Delta_{\rm h} T_{\rm b}$ is shown a function of redshift from simulations with five mass resolutions, $\mathcal{N} = 64$, 128, 256, 512, and 1024, in the Dark Ages (left panel) and Saturated Coupling (right panel) cases, where we set $x_\alpha=0$ and $\infty$, respectively. All the cases are calculated for the fiducial box density ($\bar\delta=0$) and streaming velocity ($V_{bc,i}=28~{\rm km}~{\rm s}^{-1}$).} \label{fig_ResTest}
\end{figure}


\section{Impact of the complex baryonic and structure-formation physics} \label{sec:complex_physics}

\subsection{Structure formation and gas temperature}\label{sec:structure_formation}

Gas generally cools over time during the Dark Ages and early Cosmic Dawn due to adiabatic expansion. However, the development of non-linear structures results in significant heating in overdense regions at lower redshifts, as visualized in Supplementary Figure~\ref{fig_TimeEvolutionPtl} and Supplementary Video \movieTgas. In the right panels of Supplementary Figure~\ref{fig_TS}, we present the relationship between the hydrogen number density ($n_{\rm HI}$) and gas temperature ($T_{\rm gas}$) at redshifts $z=100$, 70, 30, and 19, to illustrate the heating of baryons driven by structure formation. Note that in these panels, $T_{\rm gas}$ is labeled as $T_{\rm S}$ since they are equal for Saturated Coupling ($x_\alpha=\infty$).

The $T_{\rm gas}-n_{\rm HI}$ relation shows that $T_{\rm gas}$ increases in overdense regions due to adiabatic compression and decreases in underdense regions due to adiabatic expansion. However, $T_{\rm gas}$ does not strictly follow the adiabatic relation, $T_{\rm gas} \propto {n_{\rm HI}}^{2/3}$, due to additional processes influencing the gas temperature \cite{Barkana2005, naoz05}. At redshifts $z=100$ and 70, the overall relationship is flatter than the adiabatic expectation because Compton heating raises $T_{\rm gas}$ globally and more significantly in underdense regions, where $T_{\rm gas}$ is initially lower. As Compton heating weakens at lower redshifts, the $T_{\rm gas}-n_{\rm H}$ relation approaches the adiabatic case at $z=30$ but remains flatter in low-density regions down to $z=19$. At $z=19$, the relationship broadens considerably in overdense regions due to additional heating from shocks generated by collapsing structures, making the slope of the $T_{\rm gas}-n_{\rm HI}$ relation steeper than the adiabatic case on average.

\begin{figure}[h]
\centering
\includegraphics[width=0.8\textwidth, angle=0]{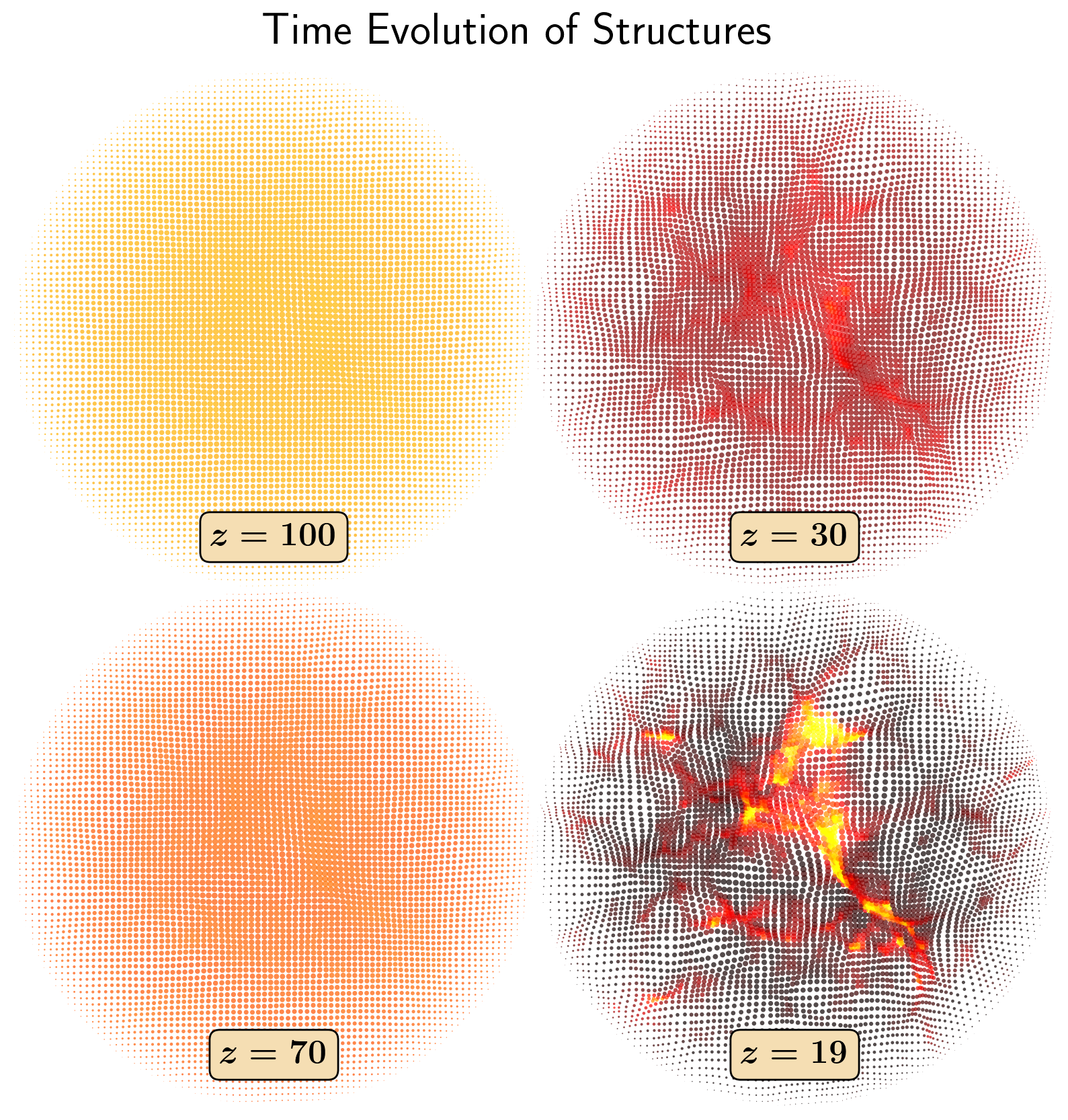} 
\caption{Gas particle distribution and temperature at $z=100,~70,~30,$ and 19 in the fiducial simulation, described in Sec.~\ref{sec:structure_formation}. Other figure details are as in Supplementary Figure~\ref{fig_Ptl}.} \label{fig_TimeEvolutionPtl}
\end{figure}

\begin{figure}[h]
\centering
\includegraphics[width=0.92\textwidth, angle=0]{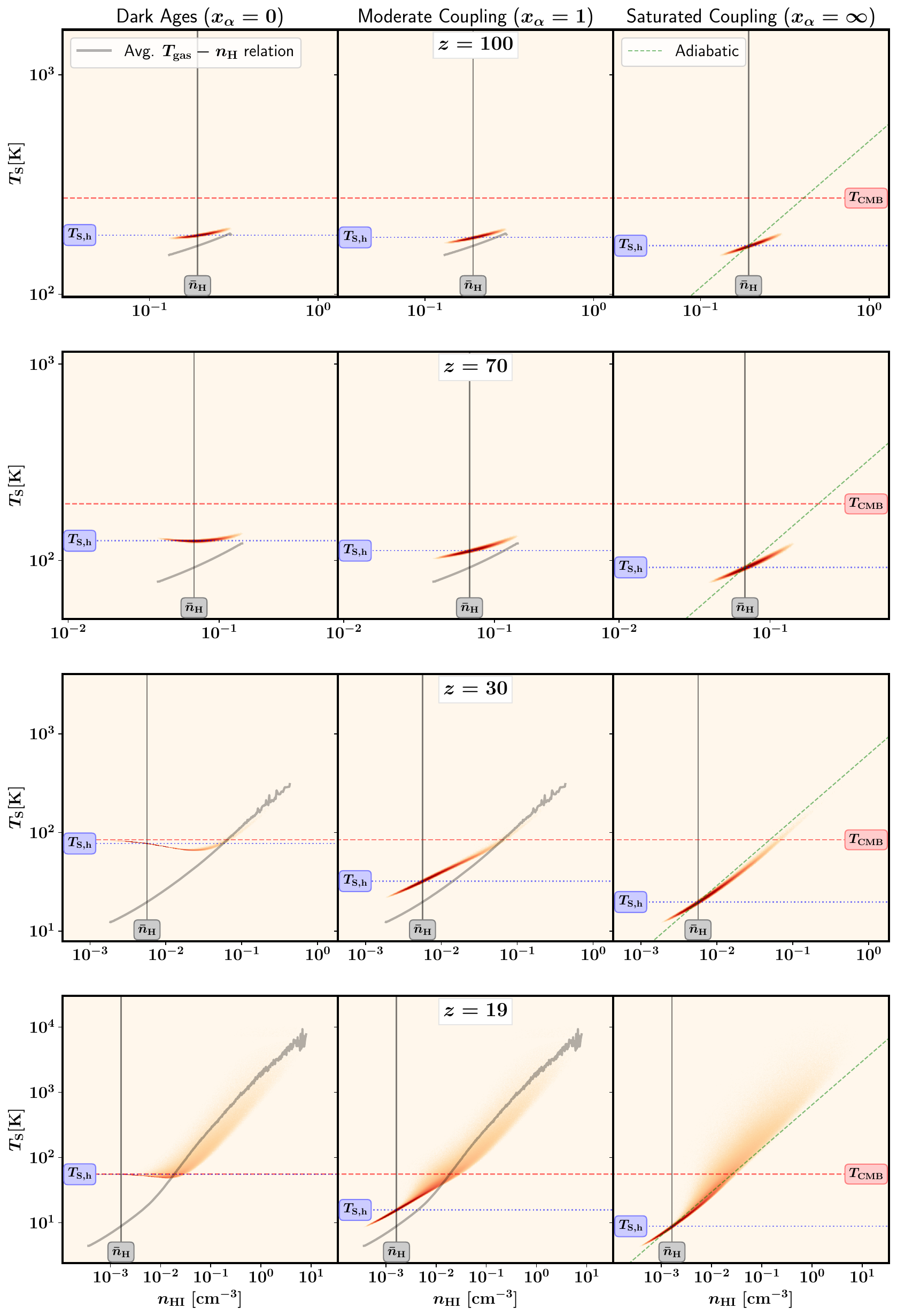} 
\caption{Scatter plot of $T_{\rm S}$ versus $n_{\rm HI}$ across different Ly$\alpha$ coupling regimes: Left, middle, and right panels show the Cosmic Dark Ages ($x_\alpha = 0$), Moderate Coupling ($x_\alpha = 1$), and Saturated Coupling ($x_\alpha = \infty$), respectively. Results are based on the simulation with $\bar{\delta} = 0$ and $V_{bc,i} = 28~{\rm km}~{\rm s}^{-1}$. Rows correspond to redshifts $z = 100$, 70, 30, and 19 (top to bottom). The red dashed line marks $T_{\rm CMB}$, and the blue dotted line marks $T_{\rm S,h}$ from the uniform simulation box. Black vertical lines indicate the cosmic mean density, $\bar{n}_{\rm HI}$. The left and middle panels show the average $T_{\rm gas}$–$n_{\rm HI}$ relation (which is based on the plots in the right panels in which $T_{\rm S} = T_{\rm gas}$) as black curves, while the diagonal green dashed lines in the right panels show the adiabatic temperature-density relation for monatomic gas, $T_{\rm gas}\propto n_{\rm HI}^{2/3}$. The full time evolution from the available redshifts is shown in Supplementary Video \movieTvsRho.} \label{fig_TS}
\end{figure}

\subsection{Collisional coupling and the Dark Ages signal}

In the upper panel of Figure~1 (main text), we show the change $\Delta_{\rm h} T_{\rm S} = T_{\rm S} - T_{\rm S,h}$ in spin temperature due to structure formation during the Cosmic Dark Ages, compared to the homogeneous Universe value, for individual SPH particles. We show results from assuming either a standard CDM scenario or a WDM-like scenario, where in the latter case we Gaussian-smoothed the ICs with $k_{\rm cut}=100~h~{\rm Mpc}^{-1}$. The case chosen for illustration is the cosmic mean density $\bar{\delta}=0$ and $V_{bc,i}=28~{\rm km}~{\rm s}^{-1}$. In the lower panel, we show the change in mean brightness temperature relative to the homogeneous Universe value, $\Delta_{\rm h} T_{\rm b} = T_{\rm b} - T_{\rm b,h}$. A time-lapse movie of the SPH particles with color-coded $\Delta_{\rm h} T_{\rm S}$ can be found in Supplementary Videos \movieDelTS~and \movieDelTSWDM~for the CDM and WDM-like cases, respectively. Additionally, we present the relationship between hydrogen number density ($n_{\rm HI}$) and $T_{\rm S}$ in the Dark Ages case for redshifts $z=100$, 70, 30, and 20 in the left panels of Supplementary Figure~\ref{fig_TS}, and the Moderate Coupling case in the middle panels. As noted before, $T_{\rm S,h}$ and $T_{\rm b,h}$ are derived in each case from the corresponding “uniform” simulation, as described in Methods.

Around $z=100$, Figure~1 shows positive $\Delta_{\rm h} T_{\rm S}$ in overdense regions (red circles) and negative $\Delta_{\rm h} T_{\rm S}$ in underdense regions (blue triangles). This occurs because $T_{\rm S}$ correlates with density due to its coupling to $T_{\rm gas}$. The Dark Ages $T_{\rm S}$-$n_{\rm HI}$ relation at $z=100$ (top left panel of Supplementary Figure~\ref{fig_TS}) is flatter in underdense regions than in overdense regions because weaker collisional coupling draws $T_{\rm S}$ relatively closer to $T_{\rm CMB}$ and further away from $T_{\rm gas}$. This asymmetry in the $T_{\rm S}$-$n_{\rm HI}$ relation leads to a positive net $\Delta_{\rm h} T_{\rm b}$, which increases over time as collisional coupling weakens. 

After $z \approx 70$, $\Delta_{\rm h} T_{\rm S}$ becomes positive in underdense regions (red triangles), where the weakening coupling drives the gas towards the high CMB temperature, while $\Delta_{\rm h} T_{\rm S}$ remains positive in overdense regions (red circles), where the coupling is stronger but $T_{\rm gas}$ is higher as well. The result is a ``U"-shaped $T_{\rm S}$-$n_{\rm HI}$ relation (mid-upper left panel of Supplementary Figure~\ref{fig_TS}) down to $z \approx 60$. The absence of negative $\Delta_{\rm h} T_{\rm S}$ causes $\Delta_{\rm h} T_{\rm b}$ to peak at approximately 0.3 mK from $z \approx 70$ to 60. The anti-correlation between $T_{\rm S}$ and $n_{\rm HI}$ gradually extends to overdensities over time. Due to the different evolution of the $T_{\rm S}$-$n_{\rm HI}$ relation in underdensities and overdensities, $\Delta_{\rm h} T_{\rm b}$ exhibits weak wiggles between $z\approx70$ and 60 (lower panel of Figure~1 and bottom left panel of Figure~2 in the main text).

After $z \approx 60$, overdense regions start to exhibit negative $\Delta_{\rm h} T_{\rm S}$ (blue circles), as the $T_{\rm S}$-$n_{\rm HI}$ relation becomes anti-correlated around the mean density. The reversal of the $T_{\rm S}$-$n_{\rm HI}$ correlation is visualized in Supplementary Video \movieDelTS. During this epoch, $T_{\rm S}$ is increasingly driven by collisional coupling, pulling $T_{\rm S}$ closer to $T_{\rm gas}$ in overdensities and less so in underdensities. As time progresses, the negative contribution from overdense regions strengthens, driving $\Delta_{\rm h} T_{\rm b}$ down and resulting in negative values at $z \lesssim 45$.

At $z \approx 27$, $\Delta_{\rm h} T_{\rm b}$ reaches a minimum of approximately -0.45 mK. Around this epoch, $T_{\rm S}$ starts exceeding $T_{\rm CMB}$ at the high-density end due to strong adiabatic and shock heating raising $T_{\rm gas}$, with which $T_{\rm S}$ couples (mid-lower left and bottom left panels of Supplementary Figure~\ref{fig_TS}). The positive signal from highly overdense regions ($n_{\rm HI}/\bar{n}_{\rm HI}\gtrsim10$) grows over time, while the negative signal from mildly overdense regions ($1\lesssim n_{\rm HI}/\bar{n}_{\rm HI}\lesssim10$) diminishes as $T_{\rm S}$ converges towards $T_{\rm CMB}$ due to weakening collisional coupling. As a result of these changes, $\Delta_{\rm h} T_{\rm b}$ rises again after $z\approx27$.

$\Delta_{\rm h} T_{\rm b}$ is suppressed by approximately 50\% in the WDM-like scenario with $k_{\rm cut}=100~h~{\rm Mpc}^{-1}$, as the smoothed ICs lead to significantly reduced structure formation at sub-Mpc scales, as illustrated in the upper panel of Figure~1 (see also Supplementary Video \movieDelTSWDM). Therefore, $\Delta_{\rm h} T_{\rm b}$ can serve as a sensitive probe of dark matter properties, which is the main implication that we emphasize throughout this work.

\subsection{Ly$\alpha$ coupling and the Cosmic Dawn signal} \label{sec:CosmicDawn}

The results for the Dark Ages are no longer applicable once the first stars emerge and begin producing a significant WF effect at $z \sim 40-20$. To illustrate this early Cosmic Dawn regime, we calculate $T_{\rm S}$ and the subsequent $T_{\rm b}$ when setting $x_\alpha=1$ (Moderate Coupling) or $x_\alpha=\infty$ (Saturated Coupling), using $T_{\rm gas}$ from our simulations. Note that this is done in post-processing, and does not require new simulation runs. Although the actual value of $x_\alpha$ is subject to highly uncertain star-formation physics during the Cosmic Dawn, our results with $x_\alpha=0,~1,$ and $\infty$ mark the possible range of $T_{\rm b}$ during this epoch. The $T_{\rm S}-n_{\rm H}$ relation for the Moderate and Saturated Coupling cases is presented in the middle and right panels of Supplementary Figure~\ref{fig_TS}, respectively. $\Delta_{\rm h} T_{\rm b}$ is shown in Supplementary Figure~\ref{fig_Vbc} for several cases. In this subsection, we focus on the cases with $V_{bc,i} = 28~{\rm km}~{\rm s}^{-1}$ at the cosmic mean, represented by the black solid curves in Supplementary Figure~\ref{fig_Vbc}.

With Saturated Coupling, $T_{\rm S}$ (which equals $T_{\rm gas}$) has a monotonically positive correlation with density at all redshifts. Structure formation leads to a positive net $\Delta_{\rm h} T_{\rm S}$ because (1) the $T_{\rm S}-n_{\rm H}$ relation is asymmetric, being steeper at high densities, and (2) the gas density distribution skews toward high densities as structures become non-linear. $\Delta_{\rm h} T_{\rm b}$ grows steeply at $z \lesssim 40$, reaching 13~mK at $ z \approx 20 $ (right panel of Supplementary Figure~\ref{fig_Vbc}), leading to an approximately $ -10\% $ correction to the homogeneous Universe value (as was previously calculated \cite{2021ApJ...923...98X} but without the streaming velocity included here or the effect of large-scale fluctuations presented in the main text).

With Moderate Coupling, $T_{\rm b}$ behaves similarly to that with Saturated Coupling but is weaker by approximately a factor of two, reaching 6~mK at $ z \approx 20 $ (middle panel of Supplementary Figure~\ref{fig_Vbc}). In this case, the $ T_{\rm S}-n_{\rm H} $ relation is not fully coupled but is strongly drawn toward the $ T_{\rm gas}-n_{\rm H} $ relation, also showing a monotonically positive correlation.

As in the Dark Ages limit, the WDM-like cases reduce $ \Delta_{\rm h} T_{\rm S}$ by approximately a quarter, a half, and three quarters for $ k_{\rm cut} = 300,~100, $ and $ 30~h~{\rm Mpc}^{-1} $, respectively. Given that $ \Delta_{\rm h} T_{\rm S} $ is generally more than an order of magnitude stronger in these early Cosmic Dawn scenarios than in the Dark Ages, future global 21-cm signal measurements for $ 15 \lesssim z \lesssim 40 $ offer promising opportunities to constrain dark matter properties. However, the factor of $\sim$2 difference between Moderate and Saturated Coupling highlights the uncertainties stemming from the unknown star formation physics, which need to be separated out in order to place robust constraints on the dark matter properties.

\subsection{Baryon-dark matter drift motion} \label{sec_streaming}

\begin{figure*}[h]
\centering
\includegraphics[width=1\textwidth, angle=0]{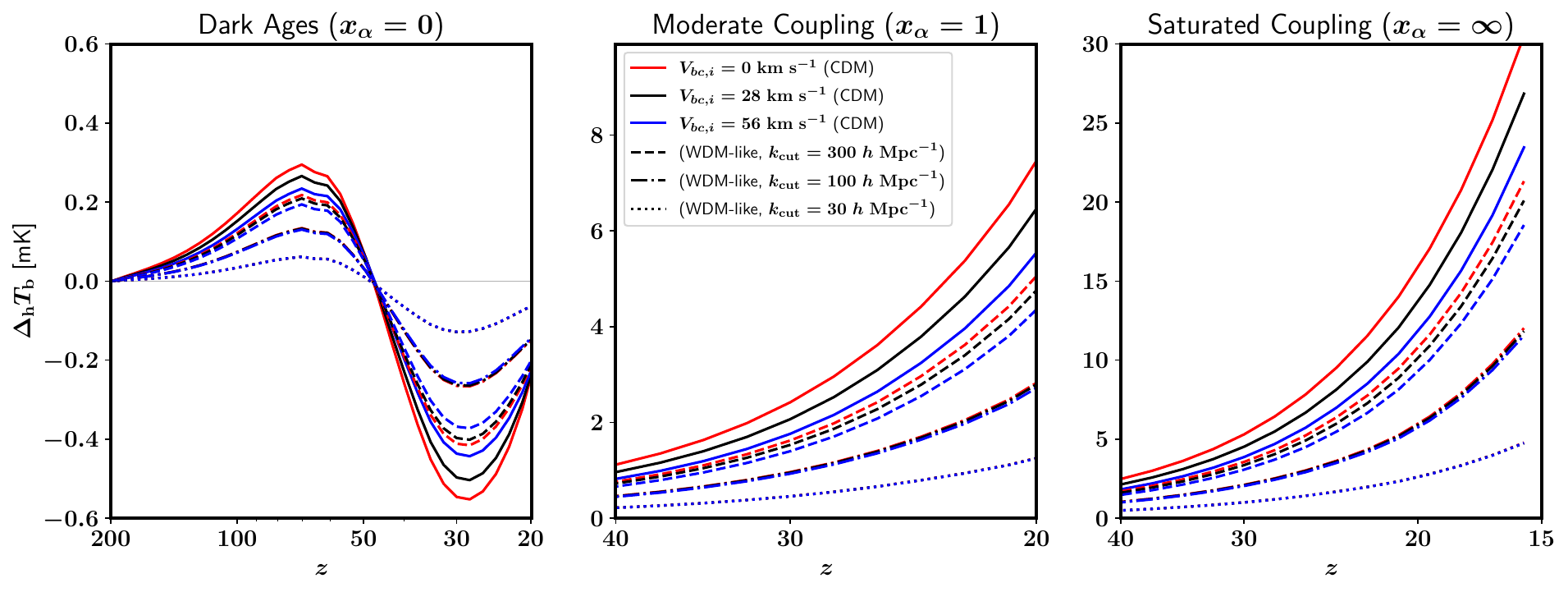} 
\caption{$\Delta_{\rm h} T_{\rm b}$ as a function of redshift at the cosmic mean density ($\bar{\delta}=0$) for $V_{bc,i}=0$, 28, and $56~{\rm km}~{\rm s}^{-1}$ and four levels of smoothing: no smoothing, $k_{\rm cut}=300$, 100, and $30~h~{\rm Mpc}^{-1}$. The WDM-like models approximately corresponds to particle masses of 25, 7, and 1.8 keV, respectively. The left, middle and right panels show the results for the Dark Ages, Moderate Coupling, and Saturated Coupling, respectively. Note the difference in the range of horizontal axis in the three panels.} \label{fig_Vbc}
\end{figure*}

Our work is the first numerical study to incorporate the baryon-dark matter streaming velocity into predicting the global 21-cm signal. Cases with different $V_{bc,i}$ values in Supplementary Figure~\ref{fig_Ptl} show how the outskirts of the main structures are smoothed out as the streaming velocity increases. Supplementary Figure~\ref{fig_Vbc} shows the impact on $\Delta_{\rm h} T_{\rm b}$ for three streaming velocities, $V_{bc,i} = 0$, 28, and $56~{\rm km}~{\rm s}^{-1}$, at four levels of smoothing in the initial conditions: no smoothing for the CDM scenario and $k_{\rm cut} = 300$, 100, and $30~h~{\rm Mpc}^{-1}$ for the WDM-like scenarios.

In the Cold Dark Matter (CDM) scenario without any smoothing, different streaming velocities produce notable variations in $\Delta_{\rm h} T_{\rm b}$. At $z=30$ in the Dark Ages case, for instance, the $\Delta_{\rm h} T_{\rm b}$ values are -0.5064, -0.4625, and -0.4059 mK for $V_{bc,i} = 0$, 28, and $56~{\rm km}{\rm s}^{-1}$, respectively, with a globally average of -0.4482 mK. The impact of averaging $\Delta_{\rm h} T_{\rm b}$ over non-zero overdensities is under 1\% at this redshift. Thus, assuming zero streaming velocity, as is common in the literature, overestimates $\Delta_{\rm h} T_{\rm b}$ by 13.0\%, while using the typical value $V_{bc,i} = 28~{\rm km}~{\rm s}^{-1}$ as an approximate attempt to account for the streaming velocity, still overestimates $\Delta_{\rm h} T_{\rm b}$ by 3.26\%.

\begin{figure*}[h]
\centering
\includegraphics[width=0.8\textwidth, angle=0]{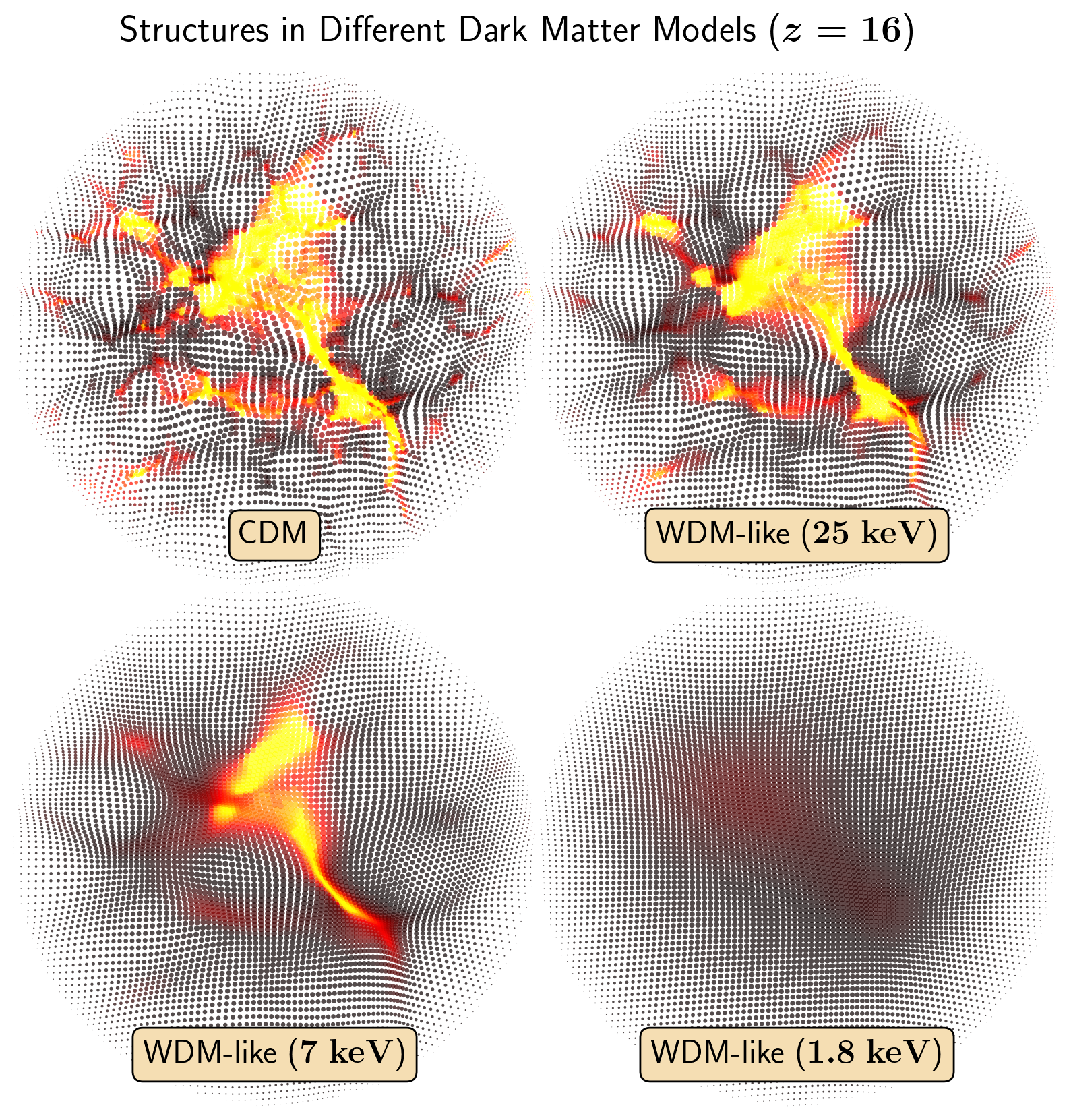} 
\caption{Gas particle distribution and temperature for different dark matter scenarios: CDM and WDM-like cases with cut-off scales $k_{\rm cut}=300,~100,$ and $30~h~{\rm Mpc}^{-1}$ at $z=16$. The WDM-like models approximately correspond to particle masses of 25, 7, and 1.8 keV, respectively. Other figure details are as in Supplementary Figure~\ref{fig_Ptl}.} \label{fig_WDMPtl}
\end{figure*}

The streaming effect is similarly significant for Moderate Coupling and Saturated Coupling. At $z=30$, $\Delta_{\rm h} T_{\rm b}$ is 2.750, 2.342, and 2.018 mK with Moderate Coupling, and 6.138, 5.216, and 4.515 mK with Saturated Coupling. The globally averaged values are 2.500 and 5.573 mK, respectively. Assuming zero streaming velocity results in a 10.0\% and 10.2\% overestimation of the signal, while $V_{bc,i} = 28~{\rm km}~{\rm s}^{-1}$ leads to a 6.32\% and 6.39\% underestimation, respectively. Note that the effect of clumping is too large in all cases that ignore the streaming velocity (i.e., set it to zero), since streaming moves the baryons away from the dark matter overdensities, reducing the gravitational force and thus the baryon clumping.

In WDM-like scenarios, the impact of the streaming motion diminishes significantly as small structures, which are most affected by streaming, are smoothed out, as illustrated in Supplementary Figure~\ref{fig_WDMPtl}. With the weakest smoothing ($k_{\rm cut} = 300~h~{\rm Mpc}^{-1}$), the globally averaged $\Delta_{\rm h} T_{\rm b}$ values are -0.3626, 1.8151, and 4.0171 mK for the Dark Ages, Moderate Coupling, and Saturated Coupling cases, respectively. Assuming zero streaming velocity gives $\Delta_{\rm h} T_{\rm b}$ values of -0.3838, 1.8394, and 4.0410 mK, giving 5.52\%, -1.32\%, and -0.594\% differences—significantly smaller effects than in the CDM scenario (13.0\%, 10.0\%, and 10.2\%). For $k_{\rm cut} = 100$ and $30~h~{\rm Mpc}^{-1}$, the differences become negligible ($\lesssim 1\%$).

\subsection{Redshift-space distortion effect} \label{sec:RSD}

As expressed in Equation~(3) in the main text, the 21-cm optical depth, $\tau$, depends on the velocity gradient in baryons, $dv_r/dr$, since the signal arises from the redshift-space distribution of hydrogen. In a homogeneous Universe, this gradient would simply equal $H(z)/(1+z)$, in terms of the (redshift-dependent) Hubble constant. However, the peculiar velocities from structure formation introduce deviations from this result on length scales where structure formation occurs, resulting in a stronger gradient in underdensities and a weaker or negative gradient in overdensities (see Supplementary Figure~\ref{fig_RSDptl} for a visualization of this effect in our simulation). This leads to a weighting bias toward higher-density regions in the global 21-cm signal. Our \texttt{21\textsc{cm}SPACE} semi-numerical grid accounts for this effect on large scales, using the linear theory result \cite{kaiser87,bharadwaj04,BLlos} which suffices at the redshifts of interest. We describe here how this effect is incorporated for non-linear sub-Mpc-scale structures within the hydrodynamical simulations.

To account for this redshift-space distortion effect, we use the redshift-space HI density for $n_{\rm HI}$, which we will denote here as $n_{{\rm HI},z}$ to distinguish it from the real-space HI density $n_{{\rm HI},r}$. For instance, mildly overdense structures appear more compressed along the line of sight (i.e., the Kaiser effect) in redshift space, effectively increasing their density compared to that in real space (i.e., $n_{{\rm HI},z}>n_{{\rm HI},r}$). We employ a novel method, not previously used in the literature, to compute $n_{{\rm HI},z}$ from our SPH simulation. Specifically, we modify the output snapshots by applying a redshift-space correction to particle locations, adopting the $z$-axis as the line of sight. We then resume the GADGET simulation with these modified snapshots to recalculate the SPH density for each particle in redshift space, outputting the results without advancing in time to avoid any unphysical evolution in the physical quantities.

\begin{figure*}[h]
\centering
\includegraphics[width=0.9\textwidth, angle=0]{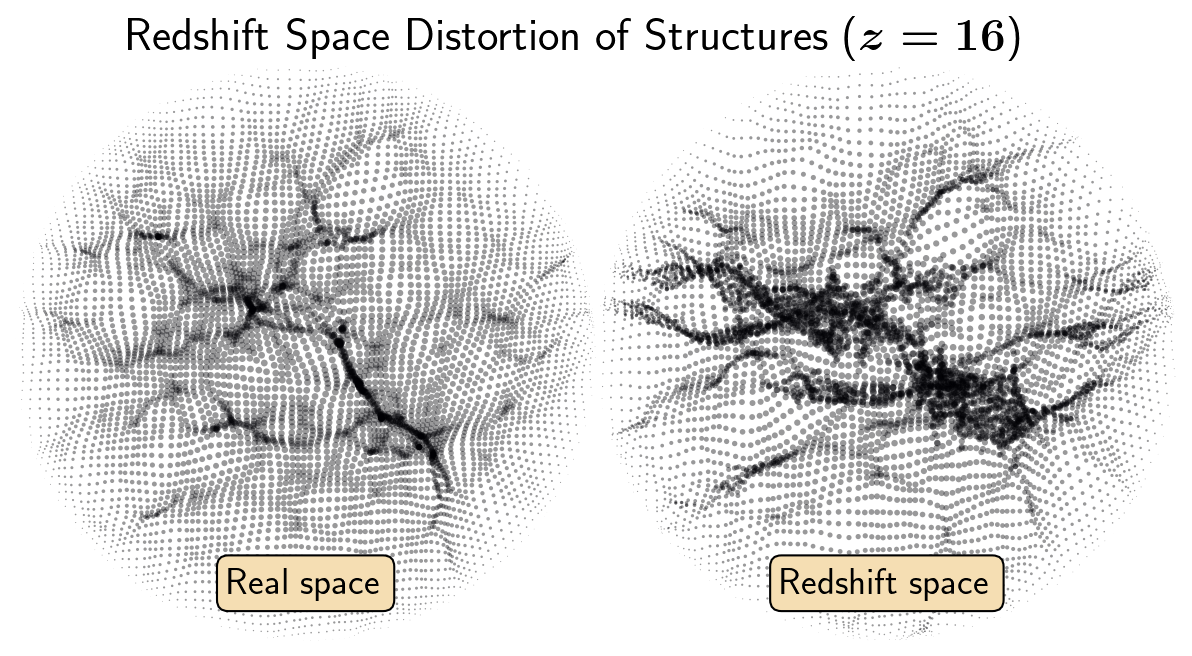} 
\caption{Gas particle distribution in real space vs.\ redshift space in the fiducial simulation at $z=16$, visualizing the redshift space distortion effect descried in Sec.~\ref{sec:RSD}. The redshift (line of sight) direction is up. Figure details are mostly the same as in Supplementary Figure~\ref{fig_Ptl}, except that only black markers are used in order to highlight differences in gas density between the two cases.} \label{fig_RSDptl}
\end{figure*}

\begin{figure*}[h]
\centering
\includegraphics[width=1\textwidth, angle=0]{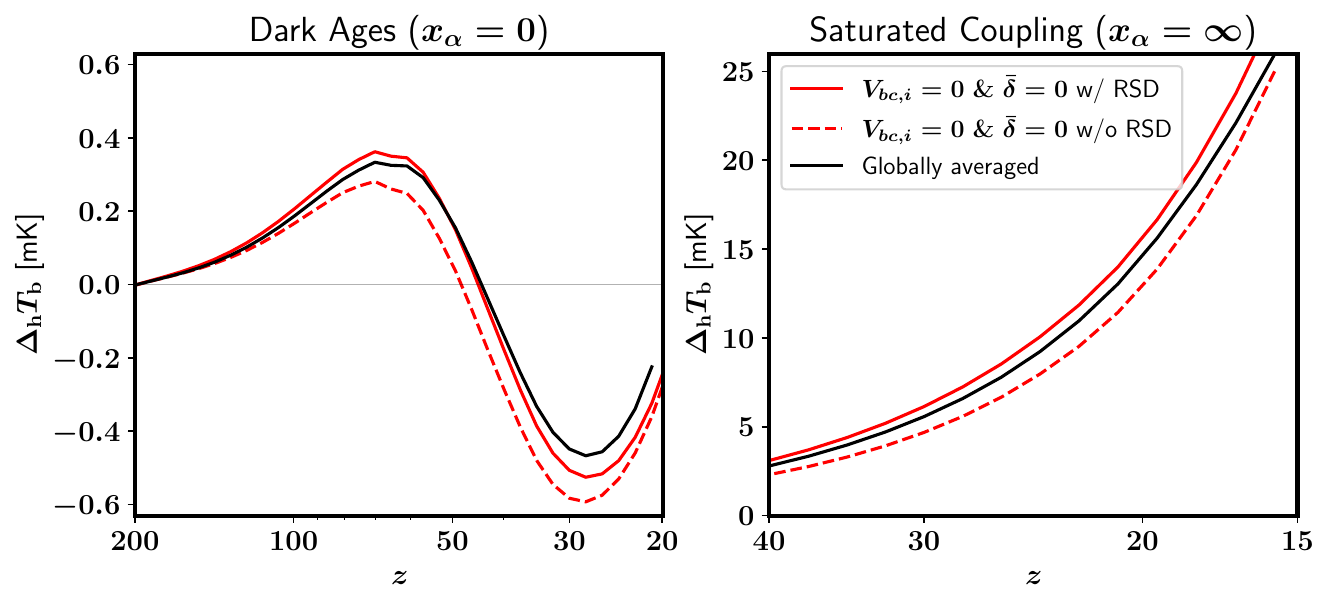} 
\caption{$\Delta_{\rm h} T_{\rm b}$ as a function of redshift with or without redshift-space distortions for $V_{bc,i}=0$ and $\bar\delta=0$, shown as red solid and dashed lines, respectively. The globally averaged $\Delta_{\rm h} T_{\rm b}$ with redshift-space distortions (just the effect of small-scale clumping) is shown as the black solid line. The left and right panels show the results in the Dark Ages and Saturated Coupling cases, respectively.} \label{fig_RSD}
\end{figure*}

Supplementary Figure~\ref{fig_RSDptl} shows how the redshift-space correction enhances particle density in overdense regions and reduces it in underdense regions. In Supplementary Figure~\ref{fig_RSD}, we show the impact of the redshift-space correction by comparing $\Delta_{\rm h} T_{\rm b}$ calculated without the correction [assuming that $n_{\rm HI}/(dv_r/dr)$ is given by $(1+z) n_{{\rm HI},r}/H(z)$] to that with the correction [assuming that $n_{\rm HI}/(dv_r/dr)$ is given by $(1+z) n_{{\rm HI},z}/H(z)$]. The redshift-space effect leads to a significant correction to $\Delta_{\rm h} T_{\rm b}$. For example, at $z=30$, $\Delta_{\rm h} T_{\rm b}$ changes by -13\% (31\%) from $-0.5825$ mK (4.683 mK) to $-0.5064$ mK (6.138 mK) in the Dark Ages (Saturated Coupling) case. We additionally show the globally averaged result with the redshift-space correction as a black solid line to illustrate how the correction compares to the effect of global averaging. Global averaging of $\Delta_{\rm h} T_{\rm b}$ changes the result by -11.5\% (-9.2\%), a smaller effect than the redshift-space correction. Notably, this work is the first to demonstrate the significance of the small-scale redshift-space effect in the global 21-cm signal.

The redshift-space correction to the net signal arises primarily from non-linear structure formation. The contribution of each particle to the net brightness temperature, $W_{V,i} \,T_{{\rm b},i}$, from Equation~(4) in the main text, scales with $\tau$ as follows:
\bea
W_{V,i} \,T_{{\rm b},i} \propto \frac{1-e^{-\tau}}{\tau},
\eea
where we have used the fact that the volume-weight, $W_{V,i}$, scales inversely with particle density and, consequently, with $\tau$ as well. In the small-$\tau$ (optically thin) limit, the right-hand side approaches a constant, indicating that the change in $\tau$ in going from real to redshift space has no effect on the net average $T_{\rm b}$. This is because the total number of 21-cm photons reaching us should remain unchanged in an optically thin Universe.

In reality, $\tau$ is typically around 0.1 near the cosmic mean density (at the high redshifts considered here), and non-linear structure formation results in $\tau\gtrsim1$ in overdense regions. Since $(1-e^{-\tau})/\tau$ decreases as $\tau$ increases, the redshift-space distortions suppress the contributions from overdense regions, where structures are mostly compressed (analogous to the Kaiser effect), while enhancing the contributions from underdense regions, where structures are stretched and $\tau$ is low. 

For most hydrogen atoms, except in very dense regions at the peaks of structures, $T_{\rm S} - T_{\rm CMB}$ is negative during early cosmic times, giving a negative contribution to $T_{\rm b}$. Consequently, redshift-space distortions produce a positive change in $T_{\rm b}$ in overdense regions and a negative change in underdense regions. As structure formation progresses into the non-linear regime, the distribution of particles becomes increasingly skewed toward higher densities. This shift causes the positive contributions from overdensities to outweigh the negative contributions from underdensities, resulting in an overall positive correction to $T_{\rm b}$, as observed in our results.

\subsection{Contribution of density fluctuations on various scales} 

We find that both the small-scale (below 3~Mpc) and large-scale (above 3~Mpc) density and velocity fields are important in order to calculate the 21-cm effect of clumping $\Delta_{\rm h} T_{\rm b}$ accurately. The full effect of structure formation, which we present in the main text, includes both small- and large-scale fluctuations, and is given by the answer from the large-scale grid (with varying mean density and streaming velocity in each pixel) measured relative to the homogeneous case. As noted in the Methods, our baseline, ``homogeneous" case uses the 3~Mpc simulation box with a uniform density at the cosmic mean and with no streaming velocity. 

To isolate the effect of small-scale clumping, we use a different comparison case. We again combine the simulations with the large-scale grid, but use the results of uniform 3-Mpc simulations with varying mean density and streaming velocity but no internal fluctuations. This includes the effect of large-scale fluctuations only, and the difference between this case and the full case (which is the same except with the addition of internal clumping) can be defined as the effect of small-scale clumping.

The overall 21-cm effect of varying $\Delta_{\rm h} T_{\rm b}$ comes mostly from small-scale clumping. Still, when compared to the effect of small-scale clumping only, we find that the overall clumping effect at $z=30$ is larger by 11.1\% in the Dark Ages case, smaller by 1.76\% for Moderate Coupling, and smaller by 1.26\% for Saturated Coupling.

\section{Additional notes on observability}
\label{sec:observe}

We expand here on our analysis methods used for the section on "Observability" within the main text. As noted there, we follow the approach in our recent papers \cite{2023NatAs...7.1025M,2024MNRAS.527.1461M}, and include the thermal noise, multiple redshift bins (with $\Delta \nu=1$~MHz), and the degeneracy that makes any unobservable signal component in the shape of the synchrotron foreground. We do not consider uncertainties associated with calibration and beam modeling as well as the effects of more complex foreground modeling. Here we further explore and discuss some of these issues.

The thermal noise in a global signal measurement is \cite{Shaver1999}
\begin{equation}
\Delta T = \frac{T_{\rm sys}}{\sqrt{\Delta \nu \, t_{\rm int}}} \ ,
\label{eq:thermal_global}
\end{equation}
where $\Delta \nu$ is the bandwidth, $t_{\rm int}$ is the integration time, and we assume that the system temperature $T_{\rm sys}$ is approximately equal to the sky brightness temperature $T_{\rm sky} = 180\times (\nu/180\,{\rm MHz})^{-2.6}$\,K \cite{2006PhR...433..181F}.

The brightness temperature of the foreground sky emission increases with redshift, and the thermal noise for the global signal is proportional to the sky brightness; thus, the relative accuracy needed for foreground removal, in order for the foreground residuals to fall below the thermal noise, is independent of redshift (for a fixed integration time and frequency bin size). This still means that the foreground must be removed to an accuracy of a part in $10^6$ depending on the frequency bin size. This is quite challenging, but the terrestrial experiments are making steady progress, and we expect the space or lunar environment to make this task significantly easier; experiments in space not only avoid the ionosphere, but also offer a potentially benign environment that is extremely dry and stable (avoiding the effect of terrestrial moisture and variable weather), plus the blocking out of terrestrial radio frequency interference (RFI), particularly on the lunar far side. 

In our analysis we account for foreground removal while fitting the global signal, at least in an optimistic scenario. Specifically, we add a free parameter $A$ to the  model that we fit to the global 21-cm data, in the shape of the synchrotron foreground, i.e., $A\, \nu^{-2.6}$. In practice, in current global signal experiments, additional polynomial terms are usually added for a more realistic foreground modeling; this is likely a pessimistic approach that assumes that the instrument (antenna plus receiver) response cannot be modeled accurately, and a general purpose polynomial must be used as an effective model, an approach that also subtracts out a substantial portion of the 21-cm signal. As global 21-cm experiments improve, there are novel ways to calibrate without having to resort to such an unsatisfactory approach (see, e.g., the REACH experiment \cite{de_lora2022}). Also, as we noted, there are reasons to hope that this will be easier in space. Even in the best-case scenario, however, a signal component of precisely the same shape as the foreground cannot be distinguished from it, and more generally, our analysis accounts for the partial degeneracy between the signal shape and that of the foreground. 

The impact of this degeneracy is quite significant, which we illustrate here for some of our main results. For the Dark Ages case and a range of $z=200-20$, in the absence of this degeneracy a global antenna could detect the CDM signal at 10.5$\sigma$ (compared to 4.81$\sigma$ with the degeneracy included, as in our results in the main text), while the effect of clumping (i.e., the difference between CDM and a homogeneous universe) could be detected with a global signal array at 10.0$\sigma$ (compared to 7.47$\sigma$ with the degeneracy). For saturated coupling over the range $z=30-20$, a global antenna could (without the degeneracy) detect the CDM signal at 455$\sigma$ (compared to 152$\sigma$ with the degeneracy) and the effect of clumping at 31.3$\sigma$ (compared to 15.4$\sigma$ with the degeneracy). In general, the expected signal changes smoothly and monotonically with redshift, as does the foreground, so there is a significant degeneracy between them. The only exception is the effect of clumping in the Dark Ages case, which has a more complicated shape with redshift, and is therefore less strongly degenerate. Still, the effect of the foreground degeneracy is significant in all cases. 

We also consider the effect of a possibly more complex foreground that requires additional free parameters in order to model it. As noted above, in the main text we assumed a single free parameter. We now consider that foreground model as the beginning of a polynomial in frequency, and add one or two additional terms, each with a spectral shape that is $\nu$ times the previous one, with an amplitude that is an additional free parameter.

For the Dark Ages case and a range of $z=200-20$, a global antenna could detect the CDM signal at 4.8$\sigma$, 4.8$\sigma$, or 1.8$\sigma$, with 1, 2, or 3 foreground parameters, while the effect of clumping could be detected with a global signal array at 7.5$\sigma$, 3.4$\sigma$, or 3.4$\sigma$, respectively. For saturated coupling over the range $z=30-20$, a global antenna could detect the CDM signal at 150$\sigma$, 16$\sigma$, or 0.34$\sigma$, and the effect of clumping at 15$\sigma$, 3.4$\sigma$, or 0.42$\sigma$, respectively (with a global signal array giving 10 times higher significance). Thus, the effect in the Dark Ages case is significant though mild, and also discontinuous due to the various levels of degeneracy. The effect in the Saturated Coupling case is much larger due to the smoothness of the signal, but we emphasize that this is an artificial result of the fact that we have illustrated this case with the simple assumption of a fixed Saturated Coupling, while (as we illustrate here below) any realistic astrophysical model will give a rapidly rising coupling strength that will produce a rapidly varying signal with frequency.

We briefly consider also the possible degeneracy between two major parameters of interest that both similarly affect the clumping signal that we have focused on.
One is $k_{\rm cut}$, the cutoff wavenumber in the power spectrum of WDM-like models, and the other is $x_\alpha$, the strength of the Lyman-$\alpha$ coupling. As is clear from Figure~2, both parameters strongly affect the amplitude of the effect of clumping, but when modeling the signal we would aim to separate them out in order to gain information about dark matter physics that is independent of the complications of astrophysics. The top panel of Supplementary Figure~\ref{fig_degen} shows that worries about degeneracy are justified, since the logarithmic derivatives of the clumping signal with respect to the two parameters have quite similar shapes with redshift. However, this comparison is misleading. In this work we have illustrated the strength of various 21-cm signals in cases with fixed levels of Lyman-$\alpha$ coupling. In contrast, in any realistic astrophysical model, galaxy formation accelerates exponentially during cosmic dawn, resulting in a rapidly rising coupling. We illustrate this for one vanilla astrophysical model in the bottom panel of Supplementary Figure~\ref{fig_degen}; in this example, the Lyman-$\alpha$ coupling parameter rises by four orders of magnitude from redshift 40 to 20, while the total coupling parameter (which also includes the effect of collisions) varies slowly from $z=40$ to 33 and then rises by two orders of magnitude by redshift 20. On the other hand, the effect of a power spectrum cutoff should vary smoothly throughout this redshift range and correspond realistically to a constant $k_{\rm cut}$ (which in WDM and similar models is effectively set by the time of cosmic recombination and remains essentially fixed afterwards). 

\begin{figure}[h]
\centering
\includegraphics[width=.7\textwidth, angle=0]{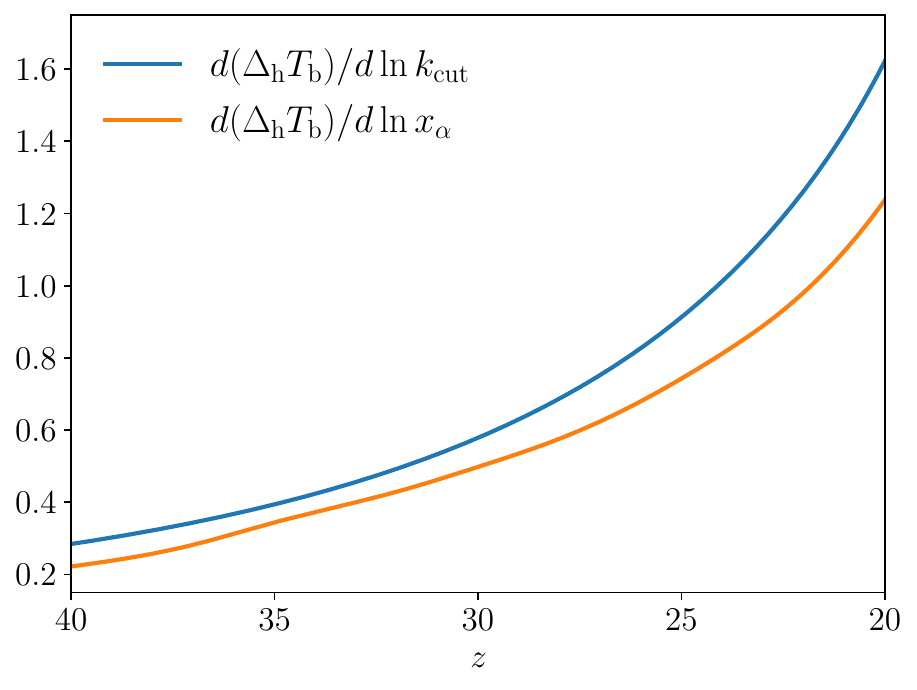}
\includegraphics[width=.7\textwidth, angle=0]{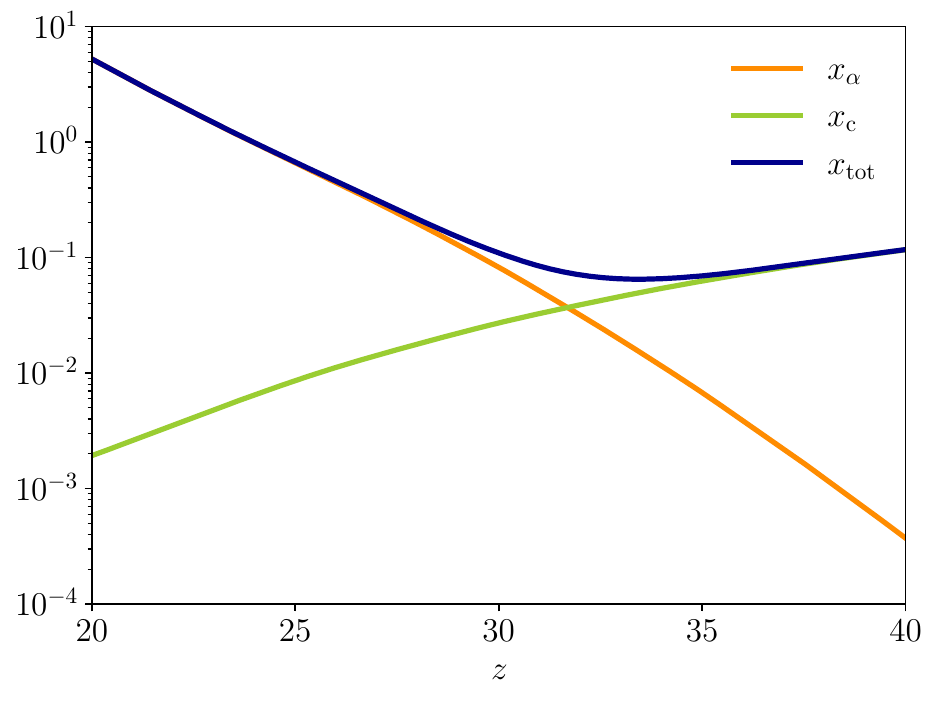}
\caption{\textbf{Top panel:} The dependence of the clumping signal (i.e., the difference between the CDM and Homogeneous cases) on logarithmic changes in two parameters: $k_{\rm cut}$, the cutoff wavenumber in the power spectrum of WDM-like models; and $x_\alpha$, the strength of the Lyman-$\alpha$ coupling. The derivatives are estimated with respect to the model with $k_{\rm cut}=100\ h$~Mpc$^{-1}$ and  $x_\alpha=1$ (Moderate Coupling). \textbf{Bottom panel:} Illustration of the dependence on redshift of the Lyman-$\alpha$ coupling parameter $x_\alpha$, the collisional
coupling parameter $x_{\rm c}$, and their sum (the total coupling parameter $x_{\rm tot}$), for a vanilla astrophysical model.} \label{fig_degen}
\end{figure}

\clearpage
\bibliography{reference}